\documentclass{article}
\usepackage{graphicx}
\usepackage{amsmath}
\usepackage{amssymb}
\usepackage{amsthm}
\usepackage{multicol}
\usepackage{multirow}
\usepackage{caption}
\usepackage{subcaption}
\usepackage{url}
\usepackage[mathscr]{euscript}
\usepackage{array}
\usepackage{float}
\usepackage{xspace}
\usepackage{tikz}
\usepackage{comment}
\usepackage{tabularx}
\usepackage{todonotes}
\usepackage{bm}
\usepackage{enumitem}
\usepackage{hyperref}
\usepackage{tikz}
\usepackage{thmtools}
\usepackage{mathrsfs}
\usepackage{thm-restate}
\usepackage{stmaryrd,nicefrac,mathtools}
\usetikzlibrary{shapes, arrows, calc, arrows.meta, fit, positioning, patterns}
\tikzset{  
    -Latex,auto,node distance =1 cm and 0.5 cm, thick,
    state/.style ={circle, fill=lightgray, draw, minimum width = 0.01 cm}, 
    emptystate/.style ={}, 
    breakarrow/.style={dashed}
    point/.style = {circle, draw, inner sep=0.18cm, fill, node contents={}}, 
    bidirected/.style={Latex-Latex,dashed}, 
    el/.style = {inner sep=2.5pt, align=right, sloped}  
} 

\newtheorem{theorem}{Theorem}[section]

\newtheorem{definition}[theorem]{Definition}
\newtheorem{lemma}[theorem]{Lemma}

\newtheorem{corollary}[theorem]{Corollary}
\newtheorem{proposition}[theorem]{Proposition}

\newtheorem{observation}[theorem]{Observation}

\newtheorem{remark}[theorem]{Remark}

\newenvironment{proofidea}{%
    \begin{proof}[Proof idea]
    \itshape
}{
    \end{proof}
}

\newcommand{\keywords}[1]{\par\addvspace{1em}\noindent\textbf{Keywords: }#1}

\begin{document}

\title{Tree Rewriting Calculi for Strictly Positive Logics}

\author{Sofía Santiago-Fernández\thanks{\url{sofia.santiago@ub.edu}} \and David Fernández-Duque\thanks{\url{fernandez-duque@ub.edu}} \and Joost J. Joosten\thanks{\url{jjoosten@ub.edu}}}

\date{}

\maketitle

\begin{abstract}
We study strictly positive logics in the language $\mathscr{L}^+$, which constructs formulas from $\top$, propositional variables, conjunction, and diamond modalities. We begin with the base system $\bf K^+$, the strictly positive fragment of polymodal $\bf K$, and examine its extensions obtained by adding axioms such as monotonicity, transitivity, and the hierarchy-sensitive interaction axiom $(\sf J)$, which governs the interplay between modalities of different strengths. The strongest of these systems is the Reflection Calculus ($\bf RC$), which corresponds to the strictly positive fragment of polymodal $\bf GLP$.

Our main contribution is a formulation of these logics as tree rewriting systems, establishing both adequacy and completeness through a correspondence between $\mathscr{L}^+$ formulas and inductively defined modal trees. We also provide a normalization of the rewriting process, which has exponential complexity when axiom $(\sf J)$ is absent; otherwise we provide a double-exponential bound. By introducing tree rewriting calculi as practical provability tools for strictly positive logics, we aim to deepen their proof-theoretic analysis and computational applications.
\end{abstract}


\section{Introduction}
Modal logics provide an attractive alternative to first or higher order logic for computational applications, largely due to the fact that they often enjoy a decidable consequence relation while remaining expressive enough to describe intricate processes.
However, decidability alone does not suffice for practical implementation when complexity comes at a hefty price tag; even propositional logic is {\sc np}-complete, which quickly becomes intractable as formula size and especially the number of variables is large.

This is no longer an issue when working in {\em strictly positive} fragments (see e.g.~\cite{kikot2019kripke}), which in contrast enjoy a polynomially decidable consequence relation.
Strictly positive formulas do not contain negation and instead are built from atoms and $\top$ using conjunction and $\Diamond$ (or, more generally, a family of modalities $\langle i \rangle$ indexed by $i$ in some set $I$).
Strictly positive formulas tend to be contingent, so validity and satisfiability are no longer the most central problems, but the consequence relation is indeed useful for example for reasoning about ontologies and is the basis for some description logics~\cite{BaaderCalvaneseEtAl2007}.

We consider strictly positive logics that extend $\bf K^+$ \cite{Beklemishev2018-BEKANO}, the basic strictly positive fragment of polymodal logic $\bf K$, by adding combinations of prominent axioms that have been considered in the literature, e.g.~monotonicity, transitivity (cf.~\cite{Thriller}), and $(\sf J)$ (cf. \cite{Beklemishev2012, dashkov2012positive}).
The logic $\bf K^+$ is a notational variant of the description logic $\mathcal EL$, useful in artificial intelligence and knowledge representation due to its balance between expressive power and computational complexity~\cite{BaaderKM99}.
Several of its proper extensions are also of interest, such as $\bf K4^+$ \cite{kikot2019kripke}, the strictly positive fragment of $\bf K4$; while the most powerful extension we consider is the {\em Reflection Calculus} ($\bf RC$)~\cite{dashkov2012positive,Beklemishev2014Positive}, the strictly positive fragment of Japaridze's polymodal provability logic $\bf GLP$~\cite{Japaridze:1988}. Beklemishev has shown how $\bf GLP$ can be used to perform a proof-theoretic analysis of Peano aritmetic and its fragments~\cite{Beklemishev:2004}; however, the logic $\bf GLP$ is notoriously difficult to work with, especially as it is not Kripke-complete. In contrast, its strictly positive fragment is rather tame from both a theoretical and computational point of view, yet suffices for the intended proof-theoretic applications. Furthermore, Dashkov \cite{dashkov2012positive} showed that the unimodal fragment of $\bf RC$ corresponds to the strictly positive fragments of both $\bf K4$ and $\bf GL$, further underscoring the versatility of $\bf RC$.

The current work is inspired by two distinct ideas that have arisen in the study of strictly positive logics. The first is the tree representation of formulas, which yield a way to decide strictly positive implications.
This was developed by Kikot et al.~\cite{kikot2019kripke} in a general setting and by Beklemishev~\cite{Beklemishev2014Positive}
in the context of $\bf RC$. In both cases, one assigns to each strictly positive formula $\varphi$ a finite, tree-like Kripke model $T(\varphi)$ with the crucial property that $\varphi\to\psi$ is valid if and only if $T(\varphi) \models \psi$. Thus the study of strictly positive fragments can be reduced to the study of their tree-like Kripke models.

The second is the connection of strictly positive calculi to term rewrite systems.
Strictly positive formulas and, particularly, those built exclusively from $\top$ and the modalities $\langle i\rangle$, may be regarded as {\em words} (or `worms'). This has prompted Beklemishev~\cite{Beklemishev2018-BEKANO} to view strictly positive fragments as term-rewriting systems~\cite{baader1998term}, but connections between such systems and modal logic are not new and can be traced back to Foret~\cite{Foret1992}.

Term rewriting is a discipline that integrates elements of logic, universal algebra, automated theorem proving, and functional programming.
It has applications in algebra (e.g. Boolean algebra), recursion
theory (computability of rewriting rules), software engineering and programming languages
(especially functional and logic programming \cite{Terese03}), 
with the $\lambda$-calculus perhaps being the most familiar example~\cite{LambdaBook}. 
Of particular interest to us, tree rewriting systems~\cite{TreePaper} are term rewriting systems (also known as abstract rewriting systems) such that terms are trees.  

Our approach is to recast the considered strictly positive logics as tree rewriting systems. When terms represent formulas, rewrite rules are similar to deep inference rules (cf. \cite{brunnler2004deep, guglielmi2015deep}), i.e.~rules which may be applied to strict subformulas of the displayed formulas.
This is the approach taken by Stewart and Stouppa~\cite{stewart2004systematic} in their systematic presentation for several unimodal logics, and by Shamkanov~\cite{shamkanov2015nested} for developing a cut-free calculus for $\bf GLP$.
As is the case for other technical differences between $\bf GLP$ and $\bf RC$, our tree rewriting calculus makes up for the loss in expressive power with increased simplicity and transparent Kripke semantics.

In the parlance of rewrite systems, cut-elimination can be viewed as a normalization procedure for derivations.
Although our setting lacks a direct analogue of the cut rule, we provide a normalization of the rewriting process. While the rewrite normalization for logics not including ($\sf J$) is achieved with only exponential complexity, when ($\sf J$) is considered we provide a double-exponential bound. Nevertheless, the normalization of the rewriting process significantly deepens our understanding of the system's dynamics, providing valuable insights into the rewriting process and the interactions among rules. Beyond their theoretical implications, they also offer a robust framework for proof search methodologies by reducing the search space: thanks to the normalization theorems, we can focus solely on normalized rewriting sequences, significantly mitigating redundancy. This reduction improves computational efficiency~\cite{goubault2001proof} and streamlines proof search. Furthermore, it serves as a practical guide for implementing the rewriting process in proof assistants~\cite{newborn2000automated}.

This work builds upon \cite{santiago2024tree}, extending it in several significant ways. In particular, we generalize the framework to define tree rewriting systems for a broader class of strictly positive logics. Moreover, we refine the techniques for reasoning about tree structures, ultimately providing specific upper bounds for the normalization theorem of the rewriting process.

In our presentation, we make use of the inductive definition of lists within the framework of type theory (cf. \cite{pierce2002types, fitting2002types}) to define the trees in our tree rewriting systems. The use of lists allows to define inductive structures with an order, facilitating the specification of internal positions and transformations for rewriting systems, and its formalization in proof assistants. Since lists play such a central role in our work, we conclude this introduction by establishing some notation. A list of elements of type $\mathscr{A}$ is either the empty list $\varnothing$ or $[x] \smallfrown L$ for $x$ an element of type $\mathscr{A}$, a list $L$ of elements of type $\mathscr{A}$ and $\smallfrown$ the operator of concatenation of lists. We write $x \smallfrown L$ and $L \smallfrown x$ to denote $[x] \smallfrown L$ and $L \smallfrown [x]$, respectively. The length of a list $L$ is denoted by $|L|$.

\section{Strictly positive logics}
We consider the language of strictly positive formulae $\mathscr{L}^+$ composed of propositional variables $p$,$q$,..., in ${\sf Prop}$, constant $\top$, and connectives $\wedge$ for conjunction and $\langle \alpha \rangle$ for diamond modalities for each $\alpha \in \mathbb{N}$. The strictly positive formulae $\varphi$ of $\mathscr{L}^+$ are generated by the following grammar: 
\begin{equation*}
\varphi ::= \top \hspace{0.1cm}|\hspace{0.1cm} p \hspace{0.1cm}|\hspace{0.1cm} \langle \alpha \rangle \varphi \hspace{0.1cm}|\hspace{0.1cm} (\varphi \wedge \varphi), \hspace{0.2cm} \alpha \in \mathbb{N} \text{ and } p \in {\sf Prop}.
\end{equation*}

\textit{Sequents} are expressions of the form $\varphi \vdash \psi$ for $\varphi, \psi \in \mathscr{L}^{+}$. The \textit{modal depth} of $\varphi$, denoted by ${\sf md}(\varphi)$, is recursively defined as ${\sf md} (\top) := 0$, ${\sf md} (p) := 0$ for $p \in {\sf Prop}$, ${\sf md} (\langle \alpha \rangle \varphi) := {\sf md} (\varphi) + 1$ and ${\sf md} (\varphi \wedge \psi) := {\sf max}\{{\sf md} (\varphi), {\sf md} (\psi)\}$. 

Polymodal $\bf K$ can be readily adapted to its strictly positive variant, where most notably the necessitation rule is replaced by distribution for each diamond modality. Hence, the basic sequent-style system $\bf K^+$ (cf. \cite{Beklemishev2018-BEKANO}) is given by the following axioms and rules:

\begin{description}
\item $\varphi \vdash \varphi$; $\varphi \vdash \top$; if $\varphi \vdash \psi$ and $\psi \vdash \phi$ then $\varphi \vdash \phi$;
\item $\varphi \wedge \psi \vdash \varphi$ and $\varphi \wedge \psi \vdash \psi$;
\item if $\varphi \vdash \psi$ and $\varphi \vdash \phi$ then $\varphi \vdash \psi \wedge \phi$;
\item if $\varphi \vdash \psi$ then $\langle \alpha \rangle \varphi \vdash \langle \alpha \rangle \psi$ (distribution).
\end{description}

The conjunction of a finite list $\Pi$ of strictly positive formulas, denoted $\bigwedge \Pi$, is defined as $\top$ for $\Pi = \varnothing$, and $\varphi \wedge \bigwedge \hat{\Pi}$ for $\Pi = \varphi \smallfrown \hat{\Pi}$. Notably, for any finite lists $\Pi_1$ and $\Pi_2$ of strictly positive formulas, $\bigwedge (\Pi_1 \smallfrown \Pi_2)$ is equivalent in $\bf K^+$ to $\bigwedge \Pi_1 \wedge \bigwedge \Pi_2$.

Aside from the basic logic $\bf K^+$, we consider the following additional standard axioms of transitivity, monotonicity, and ($\sf J$), which may be included:

\begin{description}
\item \textbf{($\sf 4$)} $\langle \alpha \rangle \langle \alpha \rangle \varphi \vdash \langle \alpha \rangle \varphi$;
\item \textbf{($\sf m$)} $\langle \alpha \rangle \varphi \vdash \langle \beta \rangle \varphi$, $\alpha > \beta$;
\item \textbf{($\sf J$)} $\langle \alpha \rangle \varphi \wedge \langle \beta \rangle \psi \vdash \langle \alpha \rangle (\varphi \wedge \langle \beta \rangle \psi)$, $\alpha > \beta$.
\end{description}

A logic ${\bf L}$ in the language $\mathscr{L}^+$ is called a \textit{strictly positive logic} ($spi$-$logic$) if it extends $\bf K^+$. The logic axiomatized over $\bf K^+$ by axiom $(\sf 4)$ is $\bf K4^+$ (cf. \cite{kikot2019kripke}), the strictly positive fragment of polymodal $\bf K4$. Extending $\bf K^+$ with axioms $(\sf 4)$, $(\sf m)$, and $(\sf J)$ results in the \textit{Reflection Calculus} ($\bf RC$ \cite{Beklemishev2012, dashkov2012positive}), which is the strictly positive fragment of Japaridze's polymodal logic $\bf GLP$.

For any spi-logic ${\bf L}$, we use ${\bf L} \cup \mathscr{A}$ to denote the logic obtained by extending ${\bf L}$ with axioms from $\mathscr{A} \subseteq \{({\sf 4}), ({\sf m}), ({\sf J})\}$. We write ${\bf L} \subseteq \mathscr{A}$ to indicate that ${\bf K}^+ \cup \mathscr{A}$ extends ${\bf L}$. The notation $\varphi \vdash_{\bf L} \psi$ means that $\varphi \vdash \psi$ is provable in ${\bf L}$, while $\varphi \equiv_{\bf L} \psi$ indicates that $\varphi \vdash_{\bf L} \psi$ and $\psi \vdash_{\bf L} \varphi$. If the context allows, we omit the subscript and write $\varphi \vdash \psi$ and $\varphi \equiv \psi$ for $\varphi \vdash_{\bf K^+} \psi$ and $\varphi \equiv_{\bf K^+} \psi$, respectively.

\section{Modal trees}

The tree rewriting calculi we present are based on a correspondence between the language of $\mathscr{L}^+$ and the inductively defined set of modal trees. 

\subsection{The structure of modal trees}

Modal trees are finite labelled trees with nodes labelled with lists of propositional variables and edges labelled with natural numbers. These trees can be thought of as tree-like Kripke models of the form $(\mathscr{W}, \{R_\alpha\}_{\alpha \in \mathbb{N}}, v)$. Hence, a natural number $\alpha$ labels an edge if $R_\alpha$ relates the corresponding nodes, while a node is labeled with the list of propositional variables that are true at that node according to the valuation $v$.

For technical reasons, particularly when presenting the rewriting calculi and formalizing our results in a proof assistant, it is more convenient to define modal trees inductively. Specifically, we arrange the children of each node in a list rather than a set. This gives us a natural order, which ensures that positions within the modal tree are clearly defined and unambiguous.

\begin{definition}[$\sf Tree^{\diamond}$] The set of {\em modal trees $\sf Tree^{\diamond}$} is defined recursively to be the set of pairs $\langle \Delta; \Gamma \rangle$, where $\Delta$ is a finite list of propositional variables and $\Gamma$ is a finite list of pairs $(\alpha, {\tt T})$, with $\alpha \in \mathbb{N}$ and ${\tt T} \in {\sf Tree^{\diamond}}$.
\end{definition} 

Elements of $\sf Tree^{\diamond}$ are denoted by ${\tt T} $ and ${\tt S}$. We use distinct notation to make it clear whether a pair represents a modal tree ($\langle \cdot ; \cdot \rangle$) or a combination of an ordinal and a modal tree ($(\cdot , \cdot)$). The \textit{root} of a modal tree $\langle \Delta ; \Gamma \rangle$ is $\Delta$ and the list of its \textit{children} is $[{\tt S} \mid (\alpha,{\tt S}) \in \Gamma]$. A \textit{leaf} is a modal tree with an empty list of children. We write $[f(\alpha,{\tt S}) \mid (\alpha,{\tt S}) \in \Gamma]$ to denote the list $[f(\alpha_1,{\tt S}_1),...,f(\alpha_n,{\tt S}_n)]$ for $\Gamma = [(\alpha_1,{\tt S}_1),...,(\alpha_n,{\tt S}_n)]$, $n \geq 0$ and $f$ a function of domain $\mathbb{N} \times {\sf Tree^{\diamond}}$. Moreover, we write $\gamma \in \Gamma$ and ${\tt T} \in \Gamma$ to denote $\gamma \in [\alpha \mid (\alpha, {\tt S}) \in \Gamma]$ and ${\tt T} \in [{\tt S} \mid (\alpha,{\tt S}) \in \Gamma]$, respectively, since the context permits a clear distinction.

In order to quantify transformations during rewriting, we present the following well-known tree metrics for a modal tree ${\tt T} = \langle \Delta ; \Gamma \rangle$. The {\em width} of a tree ${\tt T}$, denoted ${\sf w}({\tt T})$, is recursively defined as ${\sf w} (\langle \Delta ; \emptyset \rangle) := 1$ and ${\sf w} (\langle \Delta ; \Gamma \rangle) := {\sf max} (\{n\} \cup \{{\sf w}({\tt S}) \mid {\tt S} \in \Gamma\})$. The {\em height} of ${\tt T}$, denoted by ${\sf h}({\tt T})$, is inductively defined as ${\sf h}(\langle \Delta ; \varnothing \rangle) := 0$ and ${\sf h}(\langle \Delta ; \Gamma \rangle) := {\sf max}\{{\sf h}({\tt S}) \mid {\tt S} \in \Gamma\} + 1$. Finally, the {\em number of nodes} of ${\tt T}$, denoted by ${\sf n}({\tt T})$, is given by ${\sf n}(\langle \Delta ; \Gamma \rangle) := 1 + \sum\limits_{{\tt S} \in \Gamma} {\sf n}({\tt S})$.

The number of nodes of a tree is exponentially bounded in terms of its width and height.

\begin{lemma}
\label{BoundNodesByWH}
${\sf n}({\tt T}) \leq ({\sf w}({\tt T}) + 1)^{{\sf h}({\tt T})}$ for ${\tt T} \in {\sf Tree^{\diamond}}$.
\end{lemma}

\begin{proof}
By easy induction on ${\tt T}$, details left to the reader. 
\end{proof}

We define the sum of modal trees as the tree obtained by concatenating their roots and children.

\begin{definition}
The {\em sum} of modal trees ${\tt T}_1 = \langle \Delta_1; \Gamma_{1}\rangle$ and ${\tt T}_2 = \langle \Delta_2; \Gamma_2\rangle$ is defined as ${\tt T}_1 + {\tt T}_2 := \langle \Delta_1 \smallfrown \Delta_2; \Gamma_1  \smallfrown \Gamma_2 \rangle$.    
\end{definition}

More generally, for $\Lambda$ a finite list of modal trees, $\sum \Lambda$ is defined as $\langle \varnothing ; \varnothing \rangle$ if $\Lambda = \varnothing$ and ${\tt T} + \sum \hat{\Lambda}$ if $\Lambda = {\tt T} \smallfrown \hat{\Lambda}$.

To reference positions in modal trees, we use a standard numbering of nodes by strings of positive integers. The set of positions in a tree includes the root position, which is denoted by the empty string. For each node, the positions of its children are determined by appending the order of each child within the list of children.

\begin{definition}[Positions]
The {\em set of positions of a modal tree} ${\tt T} = \langle \Delta ; \Gamma \rangle$, denoted by ${\sf Pos}({\tt T}) \in \mathscr{P}(\mathbb{N}^*)$, is inductively defined as 
\begin{description}
    \item ${\sf Pos}(\langle \Delta ; \varnothing \rangle) := \{\epsilon\}$ for $\epsilon \in \mathbb{N}^*$ the empty string,
    \item ${\sf Pos}(\langle \Delta ; \Gamma \rangle) := \{ \epsilon \} \cup \bigcup\limits_{i = 1}^{n} \{i\mathbf{k} \mid \mathbf{k} \in {\sf Pos}({\tt S}_i)\}$ for $\Gamma = [(\alpha_1,{\tt S}_1), ... , (\alpha_n,{\tt S}_n)]$. 
\end{description}
\end{definition}

Using the precise position apparatus we can define derived notions like, for example, that of subtree.

\begin{definition}[Subtree]
The {\em subtree of ${\tt T} \in {\sf Tree^{\diamond}}$ at position $\mathbf{k} \in {\sf Pos}({\tt T})$}, denoted by ${\tt T}|_{\mathbf{k}}$, is inductively defined over the length of $\mathbf{k}$ as
\begin{description}
    \item ${\tt T}|_{\epsilon} := {\tt T}$,
    \item ${\tt T}|_{i\mathbf{r}} := {\tt S}_i|_\mathbf{r}$ for $1 \leq i \leq n$ such that ${\tt T} = \langle \Delta ; [(\alpha_1,{\tt S}_1), ... , (\alpha_n,{\tt S}_n)] \rangle$.
\end{description}
\end{definition}

We can now define subtree replacement based on the precise positioning.

\begin{definition}[Replacement]
Let ${\tt T},{\tt S} \in {\sf Tree^{\diamond}}$ and $\mathbf{k} \in {\sf Pos}({\tt T})$. The {\em tree obtained from ${\tt T}$ by replacing the subtree at position $\mathbf{k}$ by ${\tt S}$}, denoted by ${\tt T}[{\tt S}]_\mathbf{k}$, is inductively defined on the length of $\mathbf{k}$ as
\begin{description}
    \item ${\tt T}[{\tt S}]_\epsilon := {\tt S}$,
    \item ${\tt T}[{\tt S}]_{i\mathbf{r}} := \langle \Delta ; [(\alpha_1,{\tt S}_1),...,(\alpha_i,{\tt S}_i[{\tt S}]_\mathbf{r}),...,(\alpha_n,{\tt S}_n)]\rangle$ for $1 \leq i \leq n$ and \\${\tt T} = \langle \Delta ; [(\alpha_1,{\tt S}_1),...,(\alpha_n,{\tt S}_n)] \rangle$.
\end{description}
\end{definition}

Below, we outline key results concerning the notions of subtree and subtree replacement.

\begin{lemma}
\label{LemmaSubtreeReplacement}
Let ${\tt T},{\tt S}, \hat{\tt S} \in {\sf Tree^{\diamond}}$. For $\mathbf{k}$ and $\mathbf{r}$ belonging to the adequate position sets, we have the following properties:
\begin{description}
    \item $({\tt T}|_\mathbf{k})|_\mathbf{r} = {\tt T}|_\mathbf{kr}$ \hspace{0.2cm}\textit{(nested subtree)}; 
    \item ${\tt T} [{\tt T}|_\mathbf{k}]_\mathbf{k} = {\tt T}$ \hspace{0.2cm}\textit{(self-subtree replacement)}; 
    \item $({\tt T} [{\tt S}]_\mathbf{k})|_\mathbf{k} = {\tt S}$ \hspace{0.2cm}\textit{(subtree replacement consistency)}; 
    \item $({\tt T}[\hat{\tt S}]_\mathbf{k})[{\tt S}]_\mathbf{k} = {\tt T}[{\tt S}]_\mathbf{k}$ \hspace{0.2cm}\textit{(sequential replacement)};
    \item $({\tt T}[\hat{\tt S}]_\mathbf{k})[{\tt S}]_\mathbf{kr} = {\tt T}[\hat{\tt S}[{\tt S}]_\mathbf{r}]_\mathbf{k}$ \hspace{0.2cm}\textit{(nested replacement consistency)}.
\end{description}
\end{lemma}

\begin{proof}
Each statement is shown by induction on ${\tt T}$. The inductive steps proceed by cases on $|\mathbf{k}|$. Further details can be found in \cite{santiago2024tree}.
\end{proof}

\subsection{Relating formulas and modal trees}
\label{sec:Embeddings} 

The set of modal trees and the strictly positive formulae are related through the embeddings $\mathcal{T} : \mathscr{L}^{+} \longrightarrow {\sf Tree^{\diamond}}$ and ${\mathcal F} : {\sf Tree^{\diamond}} \longrightarrow \mathscr{L}^{+}$, which we introduce below. We show that composition ${\mathcal T} \circ {\mathcal F}$ serves as the identity over ${\sf Tree^{\diamond}}$, while ${\mathcal F} \circ {\mathcal T}$ acts as the identity on $\mathscr{L}^+$ modulo equivalence for ${\bf K^{+}}$. The embedding $\mathcal T$ draws inspiration from Beklemishev’s canonical tree representation of formulas \cite{Beklemishev2014Positive}, a combinatorial tool used to establish the polynomial-time decidability of $\bf RC$.

\begin{definition}[$\mathcal T$]
The {\em modal tree embedding} is the function ${\mathcal T} : \mathscr{L}^{+} \longrightarrow {\sf Tree^{\diamond}}$ inductively defined over $\mathscr{L}^{+}$ as 
\begin{description} 
    \item ${\mathcal T}(\top) := \langle \varnothing;\varnothing \rangle$,
    \item $\mathcal{T} (p) := \langle [p]; \varnothing\rangle$ for $p \in {\sf Prop}$,
    \item ${\mathcal T} (\langle \alpha \rangle \varphi) := \langle \varnothing; [( \alpha, {\mathcal T} (\varphi))] \rangle$ for $\varphi \in \mathscr{L}^{+}$ and $\alpha \in \mathbb{N}$,
    \item ${\mathcal T} (\varphi \wedge \psi) := {\mathcal T} (\varphi) + {\mathcal T} (\psi)$ for $\varphi, \psi \in \mathscr{L}^{+}$.
\end{description}
\end{definition}

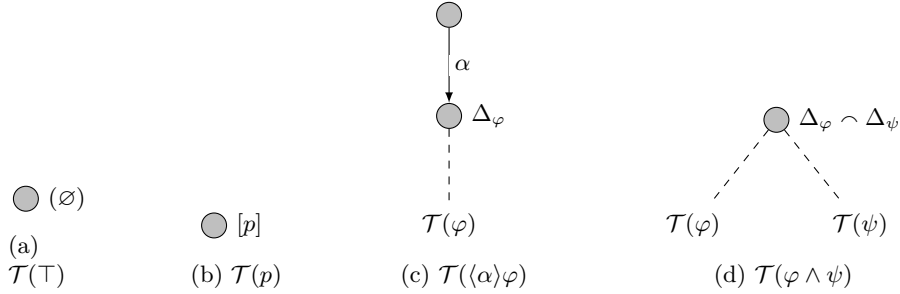
\begin{figure}[h]
\centering
    \begin{subfigure}{.1\textwidth}
    \centering
    \begin{tikzpicture}     
    \node[state][label=right:{\small ($\varnothing$)}] (a) at (0,0) {};     
    \end{tikzpicture} 
    \caption{\small ${\mathcal T} (\top)$}
    \end{subfigure}
\hfill
    \begin{subfigure}{.1\textwidth}
    \centering
    \begin{tikzpicture}     
    \node[state][label=right:{\small $[p]$}] (a) at (0,0) {};      
    \end{tikzpicture} 
    \caption{\small ${\mathcal T} (p)$}
    \end{subfigure}
\hfill
    \begin{subfigure}{.2\textwidth}
    \centering
    \begin{tikzpicture}       
    \node[state] (a) at (0,0) {};  
    \node[state][label=right:{\small $\Delta_\varphi$}] (b) [below =of a] {}; 
    \node[emptystate] (c) [below =of b] {\small ${\mathcal T} (\varphi)$};
    \draw[dashed,-] (b) edge[dashed] (c) {};
    \draw[-latex] (a) -- (b) node[fill=white,inner sep=2pt,midway] {{\small $\alpha$}};  
    \end{tikzpicture} 
    \caption{{\small ${\mathcal T} (\langle \alpha \rangle \varphi)$}}
    \end{subfigure}
\hfill
    \begin{subfigure}{.3\textwidth}
    \centering
    \begin{tikzpicture}        
    \node[state][label=right:{\small $\Delta_{\varphi} \smallfrown \Delta_{\psi}$}] (a) at (0,0) {} ;
    \node[emptystate] (b) [below left =of a] {\small ${\mathcal T} (\varphi)$};
    \node[emptystate] (c) [below right =of a] {\small ${\mathcal T} (\psi)$};
    \draw[dashed,-] (a) edge[dashed] (b) ; 
    \draw[dashed,-] (a) edge[dashed] (c) ;
    \end{tikzpicture} 
    \caption{\small ${\mathcal T} (\varphi \wedge \psi)$}
    \end{subfigure}
\caption{Modal tree embedding.}
\end{figure}

The modal depth of a formula coincides with the height of the modal tree to which the formula is mapped.

\begin{lemma}
\label{DepthMod}
${\sf h}({\mathcal T}(\varphi)) = {\sf md}(\varphi)$ for $\varphi \in \mathscr{L}^{+}$.
\end{lemma}

\begin{proofidea}
By an easy induction on $\varphi$.
\end{proofidea}

We introduce the corresponding embedding in the opposite direction.

\begin{definition}[$\mathcal F$]
The {\em strictly positive formulae embedding} is the function ${\mathcal F} : {\sf Tree^{\diamond}} \longrightarrow \mathscr{L}^{+}$ defined as 
\begin{equation*}
    {\mathcal F}(\langle \Delta ; \Gamma \rangle) := \bigwedge \Delta \wedge \bigwedge [\langle \alpha \rangle {\mathcal F}({\tt S}) \mid (\alpha,{\tt S}) \in \Gamma].
\end{equation*}
\end{definition}

For the sake of readability, we write $\Diamond \Gamma$ for $[\langle \alpha \rangle {\mathcal F}({\tt S}) \mid (\alpha,{\tt S}) \in \Gamma]$. 

The composition of the embeddings relates strictly positive formulae and modal trees as follows.

\begin{proposition}[Embedding composition]
\label{PropInvEmbedding}
    ${\mathcal T} \circ {\mathcal F} = id_{{\sf Tree^{\diamond}}}$ and ${\mathcal F} \circ {\mathcal T} = id_{\mathscr{L}^{+} / \equiv}$.
\end{proposition}

\begin{proof}
We show ${\mathcal T} \circ {\mathcal F} ({\tt T})= {\tt T}$ by induction on ${\tt T} \in {\sf Tree^{\diamond}}$. The base case is straightforward, as ${\mathcal T} ({\mathcal F} (\langle \Delta; \varnothing \rangle)) = {\mathcal T}(\bigwedge\Delta \wedge \top) = \langle \Delta; \varnothing \rangle$. For the inductive step, we see by the inductive hypothesis that
\begin{equation*}
\begin{split}
{\mathcal T}({\mathcal F}(\langle \Delta; \Gamma\rangle)) & = {\mathcal T}(\bigwedge\Delta \wedge \bigwedge \Diamond \Gamma) = {\mathcal T}(\bigwedge \Delta) + {\mathcal T} (\bigwedge \Diamond \Gamma) \\
 & = \langle \Delta; \varnothing \rangle + \sum [{\mathcal T}(\langle \alpha \rangle {\mathcal F}({\tt S})) \mid (\alpha,{\tt S}) \in \Gamma] \\
 &  = \langle \Delta;\varnothing\rangle + \sum [\langle \varnothing; [( \alpha,{\mathcal T}({\mathcal F}({\tt S})))] \rangle \mid (\alpha,{\tt S}) \in \Gamma] \\
 & = \langle \Delta ; \varnothing\rangle + \sum [\langle \varnothing; [(\alpha,{\tt S})] \rangle \mid (\alpha,{\tt S}) \in \Gamma] = \langle \Delta ; \Gamma\rangle.
\end{split}
\end{equation*}

Finally, we prove ${\mathcal F} \circ {\mathcal T} (\varphi) \equiv \varphi$ by induction on $\varphi \in \mathscr{L}^{+}$. 

\begin{enumerate}[wide, labelwidth=!, labelindent=0pt]
\item[] ${\mathcal F} \circ {\mathcal T} (\top) = \top \wedge \top \equiv \top$; and ${\mathcal F} \circ {\mathcal T} (p) = (p \wedge \top) \wedge \top \equiv p$.
\item[] If ${\mathcal F} \circ {\mathcal T} (\psi) \equiv \psi$, then ${\mathcal F} \circ {\mathcal T} (\langle \alpha \rangle \psi) = {\mathcal F} (\langle \varnothing ; [(\alpha , {\mathcal T} (\psi))] \rangle \equiv \langle \alpha \rangle {\mathcal F} \circ {\mathcal T} (\psi) \equiv \langle \alpha \rangle \psi$.
\item[] Let ${\mathcal T} (\psi) = \langle \Delta_\psi ; \Gamma_\psi \rangle$ and ${\mathcal T} (\phi) = \langle \Delta_\phi ; \Gamma_\phi \rangle$. If ${\mathcal F} \circ {\mathcal T} (\psi) \equiv \psi$ and ${\mathcal F} \circ {\mathcal T} (\phi) \equiv \phi$, then ${\mathcal F} \circ {\mathcal T} (\psi \wedge \phi) \equiv \bigwedge \Delta_\psi \wedge \bigwedge \Diamond \Gamma_\psi \wedge \bigwedge \Delta_\phi \wedge \bigwedge \Diamond \Gamma_\phi \equiv \psi \wedge \phi$.
\qedhere
\end{enumerate}
\end{proof}

\section{Tree rewriting calculi for spi-logics}

The tree rewriting calculi for strictly positive logics are defined as abstract rewriting systems for the set of modal trees.

\subsection{Preliminary definitions}

An \textit{abstract rewriting system} $\mathfrak{A}$ is a pair $(A, \{\hookrightarrow^\mu\}_{\mu \in R})$, where $A$ is a set and $\{\hookrightarrow^\mu\}_{\mu \in R}$ is a collection of binary relations on $A$, known as rewriting rules (r.r.). The reflexive transitive closure of $\hookrightarrow^\mu$ is denoted by $\hookrightarrow^{\mu *}$. Instead of writing $(a,b) \in \hookrightarrow^\mu$, we use the notation $a \hookrightarrow^\mu b$ and say that \textit{$b$ is obtained by applying $\mu$ to $a$}. When rewriting rules $\mu_1$ and $\mu_2$ are applied sequentially, we write $a \hookrightarrow^{\mu_1} \circ \hookrightarrow^{\mu_2} b$, meaning that there exists some $y \in A$ such that $a \hookrightarrow^{\mu_1} y \hookrightarrow^{\mu_2} b$. For a rewriting system $\mathfrak{A}$ and a set of additional rules $I$, we denote by $\mathfrak{A} \cup I$ the system obtained by extending the rules of $\mathfrak{A}$ with those in $I$. Given two abstract rewriting systems $\mathfrak{A}_1$ and $\mathfrak{A}_2$, we say that $\mathfrak{A}_1$ is a {\em subsystem} of $\mathfrak{A}_2$, written as $\mathfrak{A}_1 \subseteq \mathfrak{A}_2$, if there exists a set of rules $I$ such that $\mathfrak{A}_2 = \mathfrak{A}_1 \cup I$.

\subsection{Definition and structural properties}

Our tree rewriting calculi are defined with rules of the set $\mathscr{R}$, formalized in Table \ref{fig:rewrules}, which transform trees through subtree replacement. These rules fall into five \textit{kinds}: atomic, structural, replicative, decreasing, and modal. Atomic rules, $\rho^+$ (\textit{atom duplication}) and $\rho^-$ (\textit{atom elimination}), modify node labels by adding or removing propositional variables. The structural rule, $\sigma$ (\textit{child permutation}), rearranges the order of children. The replicative rule, $\pi^+$ (\textit{child duplication}), duplicates the child of a node. Decreasing rules, $\pi^-$ (\textit{child elimination}) and the $\sf 4$-rule (\textit{transitivity}), remove children and enforce transitivity. Modal rules include the $\sf m$-rule (\textit{monotonicity}), which decreases the label of an edge, and the $\sf J$-rule, which transforms the tree to simulate axiom ($\sf J$). Although some rewriting rules share names with axioms of spi-logics, we use $\mu$ to refer to the rule and $(\mu)$ for the corresponding axiom, ensuring a clear distinction.

\begin{figure}[H]
\centering  
    \begin{subfigure}{0.4\textwidth} 
        \begin{subfigure}{0.4\textwidth}
            \begin{tikzpicture}   
            \node[emptystate] (a) at (0,0) {};
            \node[state][label=left: {\small $\Delta$}] (b) [below =of a] {};
            \node[emptystate][label=right:  \hspace{0.2cm}$\hookrightarrow^{\sf 4}$] (c) [below right =of b] {};
            \node[state][label=left: {\small $\emptyset$}] (d) [below =of b] {};
            \node[emptystate] (f) [below =of d] {${\tt S}$};
            \node[emptystate] (g) [below left =of b] {};
            \draw[dashed,-] (a) edge[dashed] (b) {}; 
            \draw[dashed,-] (b) edge[dashed] (c) {};
            \draw[dashed,-] (b) edge[dashed] (g) {};
            \draw[-latex] (b) -- (d) node[fill=white,inner sep=2pt,midway] {\small $\beta$}; 
            \draw[-latex] (d) -- (f) node[fill=white,inner sep=2pt,midway] {\small $\beta$};
            \end{tikzpicture} 
        \end{subfigure}
        \hspace{0.4cm}
        \begin{subfigure}{0.4\textwidth}
            \begin{tikzpicture}   
            \node[emptystate] (a) at (0,0) {};
            \node[state][label=left: {\small $\Delta$}] (b) [below =of a] {};
            \node[emptystate] (c) [below right =of b] {};
            \node[emptystate] (d) [below =of b] {\small ${\tt S}$};
            \node[emptystate] (e) [below left =of b] {};
            \draw[dashed,-] (a) edge[dashed] (b) {}; 
            \draw[dashed,-] (b) edge[dashed] (c) {}; 
            \draw[dashed,-] (b) edge[dashed] (e) {};
            \draw[-latex] (b) -- (d) node[fill=white,inner sep=2pt,midway] {\small $\beta$}; 
            \end{tikzpicture}    
        \end{subfigure}
    \end{subfigure}
    \hspace{0.5cm}
    \begin{subfigure}{0.4\textwidth}
        \begin{subfigure}{0.4\textwidth}
            \begin{tikzpicture}  
            \node[emptystate] (a) at (0,0) {};
            \node[state][label=right: {\small $\Delta$} \hspace{0.8cm} $\hookrightarrow^{\sf J}$] (b) [below =of a] {};
            \node[state][label=left: {\small $\tilde{\Delta}$}] (c) [below left =of b] {};
            \node[emptystate] (d) [below right =of b] {\small ${\tt S}$};
            \node[emptystate] (e) [below =of c] {\small $\tilde{\Gamma}$};
            \draw[-latex] (b) -- (c) node[fill=white,inner sep=2pt,midway] {{\small $\alpha$}};
            \draw[-latex] (b) -- (d) node[fill=white,inner sep=2pt,midway] {{\small $\beta$}};
            \draw[dashed,-] (a) edge[dashed] (b);
            \draw[dashed,-] (c) edge[dashed] (e); 
            \end{tikzpicture}
        \end{subfigure}
        \hspace{0.6cm}
        \begin{subfigure}{0.3\textwidth}
            \centering
            \begin{tikzpicture} 
            \node[emptystate] (a) at (0,0) {};
            \node[state][label=right: {\small $\Delta$}] (b) [below =of a] {};
            \node[state][label=left: {\small $\tilde{\Delta}$}] (c) [below =of b] {};
            \node[emptystate](d) [below right =of c] {\small ${\tt S}$};
            \node[emptystate] (e) [below left =of c] {\small $\tilde{\Gamma}$};
            \draw[-latex] (b) -- (c) node[fill=white,inner sep=2pt,midway] {{\small $\alpha$}};
            \draw[-latex] (c) -- (d) node[fill=white,inner sep=2pt,midway] {{\small $\beta$}};
            \draw[dashed,-] (a) edge[dashed] (b);
            \draw[dashed,-] (c) edge[dashed] (e); 
            \end{tikzpicture}
        \end{subfigure} 
    \end{subfigure}
\caption{$\sf 4$-rule and $\sf J$-rule. The dash lines indicate portions of the tree that are not affected by the rule.}
\label{fig:4Jrules}
\end{figure}
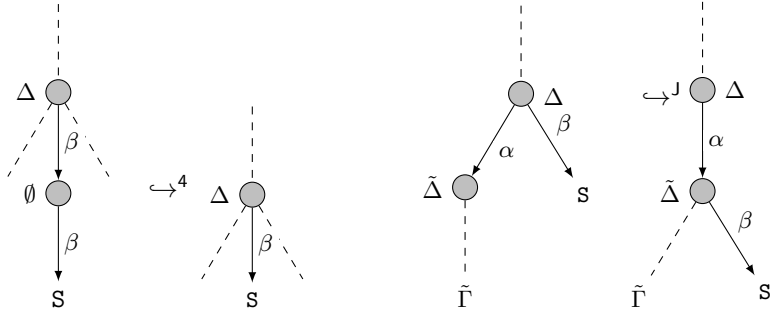

To define the rewriting rules in Table \ref{fig:rewrules}, we introduce the following notation. Let $\langle \Delta ; \Gamma \rangle \in {\sf Tree^{\diamond}}$, with $ 1 \leq i,j \leq m $ and $ n \leq |\Delta| $, where $ \Gamma = [(\alpha_1,{\tt S}_1),...,(\alpha_m,{\tt S}_m)] $. We denote the $i$-th element of $\Gamma$ as $\#_i \Gamma$. The list obtained by removing the $i$-th element from $\Gamma$ is denoted by $\Gamma^{-i}$, and the one obtained by prepending it to $\Gamma$ is denoted by $\Gamma^{+i}$. Similarly, for $\Delta$, the list obtained by removing the $n$-th element is $\Delta^{-n}$, and the one obtained by prepending it to $\Delta$ is $\Delta^{+n}$. The list obtained from $\Gamma$ by replacing its $i$-th element with $(\alpha,{\tt S})$ is written $\Gamma[(\alpha,{\tt S})]_i$. Note that we use the same notation for replacement in a list of pairs and replacement in a modal tree since the context allows for a clear distinction. Finally, the list obtained by swapping the $i$-th and $j$-th elements of $\Gamma$ is denoted by $\Gamma^{i \leftrightarrow j}$.

\renewcommand{\arraystretch}{1.5}
\begin{table}
    \centering
\begin{tabular}{| c | p{6.5cm} | c |}
\cline{1-3}
    Atomic r.r.  & \centering{${\tt T} \hookrightarrow^{\rho^+} {\tt T}[\langle \Delta^{+i} ; \Gamma \rangle]_\mathbf{k}$} & {\small $\rho^+$-rule} \\
    \cline{2-3}
    $0 < i \leq |\Delta|$ & \centering{${\tt T} \hookrightarrow^{\rho^-} {\tt T}[\langle \Delta^{-i} ; \Gamma \rangle]_\mathbf{k}$} & {\small $\rho^-$-rule} \\
\cline{1-3}
    Structural r.r. & \hfil\multirow{3}{*}{\centering ${\tt T} \hookrightarrow^{\sigma} {\tt T}[\langle \Delta ; \Gamma^{i \leftrightarrow j} \rangle]_\mathbf{k}$} & \multirow{3}{*}{{\small $\sigma$-rule}} \\
    $0 < i,j \leq |\Gamma|$ & & \\
    $i \neq j$ & & \\
\cline{1-3}
    Replicative r.r. & \hfil\multirow{2}{*}{\centering{${\tt T} \hookrightarrow^{\pi^+} {\tt T}[\langle \Delta ; \Gamma^{+i} \rangle]_\mathbf{k}$}} & \multirow{2}{*}{{\small $\pi^+$-rule}} \\
    $0 < i \leq |\Gamma|$ & & \\
\cline{1-3}
    \multirow{3}{*}{\shortstack[l]{Decreasing r.r. \\[5pt] \hfil $0 < i \leq |\Gamma|$}}  & \centering{${\tt T} \hookrightarrow^{\pi^-} {\tt T}[\langle \Delta ; \Gamma^{-i} \rangle]_\mathbf{k}$} & {\small $\pi^-$-rule} \\
    \cline{2-3}
    & \centering{${\tt T} \hookrightarrow^{{\sf 4}} {\tt T}[\langle \Delta ; \Gamma[(\beta, {\tt S})]_i \rangle]_\mathbf{k}$} & \multirow{2}{*}{{\small ${\sf 4}$-rule}} \\
    & \centering{for $\#_i \Gamma = (\beta, \langle \emptyset ; [(\beta, {\tt S})] \rangle)$} & \\
\cline{1-3}
    Modal r.r.  & \centering{${\tt T} \hookrightarrow^{{\sf m}} {\tt T}[\langle \Delta; \Gamma[(\beta,{\tt S})]_i \rangle]_\mathbf{k}$} & \multirow{2}{*}{{\small ${\sf m}$-rule}} \\
    $\alpha > \beta$ & \centering{for $\#_i \Gamma = (\alpha , {\tt S})$} & \\
    \cline{2-3}
    $0 < i,j \leq |\Gamma|$ & \centering{${\tt T} \hookrightarrow^{{\sf J}} {\tt T}[\langle \Delta ; (\Gamma[(\alpha, \langle \tilde{\Delta} ; \tilde{\Gamma} \smallfrown (\beta , {\tt S}) \rangle)]_i)^{-j} \rangle]_\mathbf{k}$} & \multirow{2}{*}{\small ${\sf J}$-rule} \\
    $i \neq j$ & \centering{for $\#_i \Gamma = (\alpha , \langle \tilde{\Delta} ; \tilde{\Gamma} \rangle)$ and $\#_j \Gamma = (\beta, {\tt S})$} & \\
\cline{1-3}
\end{tabular}
    \caption{Rules of $\mathscr{R}$ for ${\tt T} \in {\sf Tree^\diamond}$, $\mathbf{k} \in {\sf Pos}({\tt T})$ and ${\tt T}|_\mathbf{k} = \langle \Delta ; \Gamma \rangle$.}
  \label{fig:rewrules}
\end{table}

The tree rewriting calculus for $\sf K^+$ consists of atomic rules along with rules for child permutation, duplication, and elimination.

\begin{definition}[${\sf T}{\bf K^+}$] 
The {\em basic tree rewriting calculus for $\bf K^+$}, denoted ${\sf T}{\bf K^+}$, is the abstract rewriting system $({\sf Tree^{\diamond}},\{\hookrightarrow^\mu\}_{\mu \in \mathscr{K}})$ for the set $\mathscr{K} = \{\rho^+,\rho^-,\sigma, \pi^+, \pi^-\}$ of rules defined in Table \ref{fig:rewrules}.
\end{definition}

The tree rewriting calculi for the considered extensions of $\bf K^+$ are defined as extensions of ${\sf T}{\bf K^+}$.

\begin{definition}[${\sf T}{\bf L}$]
Let ${\bf L}$ be a spi-logic such that ${\bf L} \subseteq \{({\sf 4}), ({\sf m}), ({\sf J})\}$. The {\em tree rewriting calculus for ${\bf L}$}, denoted ${\sf T}{\bf L}$, is the abstract rewriting system ${\sf T}{\bf K^+} \cup \mathscr{R}_\mathscr{A}$, where $\mathscr{R}_\mathscr{A}$ is the set $\{\mu \mid (\mu) \in \mathscr{A} \text{, } {\bf L} = {\bf K^+} \cup \mathscr{A}\}$ of rules in Table \ref{fig:rewrules}.
\end{definition}

Specifically, the tree rewriting calculi for the following standard logics can be ordered using the subsystems relation.

\begin{remark}
${\sf T}{\bf K^+} \subseteq {\sf T}{\bf K4^+} \subseteq {\sf T}{\bf RC}$.
\end{remark}

Let ${\bf L}$ be a spi-logic such that ${\bf L} \subseteq \{({\sf 4}), ({\sf m}), ({\sf J})\}$. The union of the rewriting rules of a system is referred to as the \textit{rewriting relation}, denoted by $\hookrightarrow_{{\sf T}{\bf L}}$. We say that {\em the step in} ${\tt T} \hookrightarrow_{{\sf T}{\bf L}} {\tt S}$ {\em has been performed at position $\mathbf{k}$} if the applied rule replaces the subtree at position $\mathbf{k} \in {\sf Pos}({\tt T})$. The reflexive transitive closure of $\hookrightarrow_{{\sf T}{\bf L}}$ is denoted by $\hookrightarrow_{{\sf T}{\bf L}}^{*}$, and we say ${\tt T}$ {\em rewrites to} ${\tt S}$ in ${{\sf T}{\bf L}}$ if ${\tt T} \hookrightarrow^{*}_{{\sf T}{\bf L}} {\tt S}$. We say that the trees ${\tt T}$ and ${\tt S}$ are {\em ${\sf T}{\bf L}$-equivalent}, written as ${\tt T} \overset{*}{\leftrightarrow}_{{\sf T}{\bf L}} {\tt S}$, if ${\tt T} \hookrightarrow^{*}_{{\sf T}{\bf L}} {\tt S}$ and ${\tt S} \hookrightarrow^{*}_{{\sf T}{\bf L}} {\tt T}$. For simplicity, we use $\hookrightarrow$ and $\overset{*}{\leftrightarrow}$ to denote $\hookrightarrow_{{\sf T}{\bf K^+}}$ and $\overset{*}{\leftrightarrow}_{{\sf T}{\bf K^+}}$, respectively. 

Additionally, we can use a list of rewriting rules to define a sequence of rule applications. For a list $\Omega$ of rewriting rules, we define ${\tt T} \hookrightarrow^{\Omega} {\tt S}$ inductively as ${\tt T} \hookrightarrow^{*} {\tt T}$ by applying no rule if $\Omega = \varnothing$, and ${\tt T} \hookrightarrow^{\mu} \circ \hookrightarrow^{\hat{\Omega}} {\tt S}$ if $\Omega = \mu \smallfrown \hat{\Omega}$. The composition of lists of rewriting rules $\Omega_1$ and $\Omega_2$ is written ${\tt T} \hookrightarrow^{\Omega_1} \circ \hookrightarrow^{\Omega_2} {\tt S}$, and it denotes that there exists $\hat{\tt S} \in {\sf Tree^{\diamond}}$ such that ${\tt T} \hookrightarrow^{\Omega_1} \hat{\tt S} \hookrightarrow^{\Omega_2} {\tt S}$.

Below, we present several properties of the tree rewriting calculi. First, we see that modal trees with permuted lists labelling the nodes are ${\sf T}{\bf K^+}$-equivalent.

\begin{lemma}
\label{PermutationNode}
$\langle \Delta \smallfrown \Delta' ; \Gamma \rangle \overset{*}{\leftrightarrow} \langle \Delta' \smallfrown \Delta ; \Gamma \rangle$ for $\langle \Delta \smallfrown \Delta' ; \Gamma \rangle, \langle \Delta' \smallfrown \Delta ; \Gamma \rangle \in {\sf Tree^{\diamond}}$.
\end{lemma}

\begin{proof}
It suffices to show $\langle \Delta \smallfrown \Delta' ; \Gamma \rangle \hookrightarrow^{*} \langle \Delta' \smallfrown \Delta ; \Gamma \rangle$. We proceed by induction on $|\Delta'|$, making use of the associativity of the concatenation of lists and applying the atomic rules to prove the inductive step. Further details are provided in \cite{santiago2024tree}.
\end{proof}

The following lemma outlines key properties of the sum of modal trees.

\begin{lemma}
\label{LemmaRewSum}
Let ${\bf L}$ be a spi-logic such that ${\bf L} \subseteq \{({\sf 4}), ({\sf m}), ({\sf J})\}$. For ${\tt T}, {\tt T}', {\tt S}, {\tt S}' \in {\sf Tree^\diamond}$, the following hold:
\begin{enumerate}
    \item ${\tt T} \overset{*}{\leftrightarrow} {\tt T} + {\tt T}$;
    \item ${\tt T} + {\tt T}' \overset{*}{\leftrightarrow} {\tt T}' + {\tt T}$;
    \item ${\tt T} + {\tt T}' \hookrightarrow^{*} {\tt T}$ and ${\tt T} + {\tt T}' \hookrightarrow^{*} {\tt T}'$;
    \item If ${\tt T} \hookrightarrow_{{\sf T}{\bf L}}^{*} {\tt S}$, then ${\tt T} + {\tt T}' \hookrightarrow_{{\sf T}{\bf L}}^{*} {\tt S} + {\tt T}'$;
    \item If ${\tt T} \hookrightarrow_{{\sf T}{\bf L}}^{*} {\tt S}$ and ${\tt T} \hookrightarrow_{{\sf T}{\bf L}}^{*} {\tt S}'$ then ${\tt T} \hookrightarrow_{{\sf T}{\bf L}}^{*} {\tt S} + {\tt S}'$;
    \item If ${\tt T} \hookrightarrow_{{\sf T}{\bf L}}^{*} {\tt S}$ and ${\tt T}' \hookrightarrow_{{\sf T}{\bf L}}^{*} {\tt S}'$, then ${\tt T} + {\tt T}' \hookrightarrow_{{\sf T}{\bf L}}^{*} {\tt S} + {\tt S}'$.
\end{enumerate}
\end{lemma}

\begin{proof}
Below, we outline the key ideas behind each proof, leaving the finer details to the reader.

\begin{enumerate}[wide, labelwidth=!, labelindent=0pt, itemsep=0pt]
    \item ${\tt T} \hookrightarrow^{*} {\tt T} + {\tt T}$ holds by atom and child duplication; and ${\tt T} + {\tt T} \hookrightarrow^{*} {\tt T}$ by atom and child elimination.
    \item By Lemma \ref{PermutationNode} and child permutation.
    \item By atom and child elimination.
    \item By an easy induction on the number of rewriting steps performed in ${\tt T} \hookrightarrow^{*} {\tt S}$ and by cases on the rewriting rules.
    \item By (4) using (1) and (2): ${\tt T} \hookrightarrow^{*} {\tt T} + {\tt T} \hookrightarrow_{{\sf T}{\bf L}}^{*} {\tt S} + {\tt T} \hookrightarrow^{*} {\tt T} + {\tt S} \hookrightarrow_{{\sf T}{\bf L}}^{*} {\tt S}' + {\tt S} \hookrightarrow^{*} {\tt S} + {\tt S}'$. 
    \item By (5), it suffices to show that ${\tt T} + {\tt T}' \hookrightarrow_{{\sf T}{\bf L}}^{*} {\tt S}$ and ${\tt T} + {\tt T}' \hookrightarrow_{{\sf T}{\bf L}}^{*} {\tt S}'$, which follows by (3) and the hypotheses.
    \qedhere
\end{enumerate}
\end{proof}

The positioning framework of modal trees enables consistent subtree rewriting, as stated in the following result. Hence, one can effectively transform complex tree structures by rewriting subtrees, while preserving the rest.

\begin{proposition}[Deep Rewriting Property]
\label{InsideRewriting} 
Let ${\bf L}$ be a spi-logic such that ${\bf L} \subseteq \{({\sf 4}), ({\sf m}), ({\sf J})\}$. If ${\tt S} \hookrightarrow^{*}_{{\sf T}{\bf L}} {\tt S'}$, then ${\tt T} [{\tt S}]_\mathbf{k} \hookrightarrow^{*}_{{\sf T}{\bf L}} {\tt T} [{\tt S'}]_\mathbf{k}$ for ${\tt T}, {\tt S}, {\tt S}' \in {\sf Treee^\diamond}$ and $\mathbf{k} \in {\sf Pos}({\tt T})$.
\end{proposition}

\begin{proof}
We proceed by induction on the number of rewriting steps in ${\tt S} \hookrightarrow^{*}_{{\sf T}{\bf L}} {\tt S'}$. If no rewriting step is performed, then ${\tt S} = {\tt S'}$, and the result holds trivially. Otherwise, suppose that ${\tt S} \hookrightarrow^{*}_{{\sf T}{\bf L}} \hat{\tt S}$ in $n$ steps, followed by $\hat{\tt S} \hookrightarrow^{\mu} {\tt S'}$ for $\mu$ a rule of ${\sf T}{\bf L}$ applied at position $\mathbf{r} \in {\sf Pos}(\hat{\tt S})$. Since rewriting is defined via subtree replacement, we can express ${\tt S'} = \hat{\tt S}[\tilde{\tt S}]_\mathbf{r}$ for some $\tilde{\tt S} \in {\sf Tree^\diamond}$. Then, by the inductive hypothesis, we already have ${\tt T}[{\tt S}]_\mathbf{k} \hookrightarrow^{*}_{{\sf T}{\bf L}} {\tt T}[\hat{\tt S}]_\mathbf{k}$, so it remains to show ${\tt T}[\hat{\tt S}]_\mathbf{k} \hookrightarrow^{\mu} {\tt T}[\hat{\tt S}[\tilde{\tt S}]_\mathbf{r}]_\mathbf{k}$. Using Lemma \ref{LemmaSubtreeReplacement}, we have that ${\tt T}[\hat{\tt S}[\tilde{\tt S}]_\mathbf{r}]_\mathbf{k} = ({\tt T}[\hat{\tt S}]_\mathbf{k})[\tilde{\tt S}]_{\mathbf{kr}}$.
Thus, it suffices to show that 
\begin{equation*}
\text{if } \hat{\tt S} \hookrightarrow^{\mu} \hat{\tt S}[\tilde{\tt S}]_\mathbf{r}, \text{ then }{\tt T}[\hat{\tt S}]_\mathbf{k} \hookrightarrow^{\mu} ({\tt T}[\hat{\tt S}]_\mathbf{k})[\tilde{\tt S}]_\mathbf{kr}.   
\end{equation*} 
We continue by induction on the structure of ${\tt T}$.
\begin{enumerate}[wide, labelwidth=!, labelindent=0pt]
\item[\emph{Base case}]: Let ${\tt T} = \langle \Delta ; \emptyset \rangle$. Here, $\mathbf{k}$ can only be $\epsilon$, so we trivially conclude ${\tt T}[\hat{\tt S}]_\epsilon = \hat{\tt S} \hookrightarrow^{\mu} \hat{\tt S}[\tilde{\tt S}]_\mathbf{r} = ({\tt T}[\hat{\tt S}]_\epsilon)[\tilde{\tt S}]_{\epsilon \mathbf{r}}$.
\item[\emph{Inductive step}]: Let ${\tt T} = \langle \Delta ; [(\alpha_1,{\tt S}_1),...,(\alpha_m,{\tt S}_m)] \rangle$. We proceed by cases on $|\mathbf{k}|$, with the base case being trivially satisfied. For $\mathbf{k} = i\mathbf{l}$, we conclude by applying $\mu$ at position $i\mathbf{l}\mathbf{r}$ using the inductive hypothesis for ${\tt S}_i$:
\begin{equation*}
\begin{split}
{\tt T}[\hat{\tt S}]_{i\mathbf{l}} & = \langle \Delta ; [(\alpha_1,{\tt S}_1),..., (\alpha_i,{\tt S}_i[\hat{\tt S}]_{\mathbf{l}}),...,(\alpha_m,{\tt S}_m)] \rangle \\
 & \hookrightarrow^{\mu} \langle \Delta ; [(\alpha_1,{\tt S}_1),..., (\alpha_i,({\tt S}_i[\hat{\tt S}]_{\mathbf{l}})[\tilde{\tt S}]_{\mathbf{lr}}),...,(\alpha_m,{\tt S}_m)] \rangle = ({\tt T}[\hat{\tt S}]_{i\mathbf{l}})[\tilde{\tt S}]_{i\mathbf{lr}}.
\end{split}
\end{equation*}
\qedhere
\end{enumerate}
\end{proof}

\begin{corollary}
\label{InsideRewritingRepl}
Let ${\bf L}$ be a spi-logic such that ${\bf L} \subseteq \{({\sf 4}), ({\sf m}), ({\sf J})\}$. If ${\tt T}|_\mathbf{k} \hookrightarrow^{*}_{{\sf T}{\bf L}} {\tt S}$, then ${\tt T} \hookrightarrow^{*}_{{\sf T}{\bf L}} {\tt T}[{\tt S}]_\mathbf{k}$ for ${\tt T}, {\tt S} \in {\sf Tree^\diamond}$ and $\mathbf{k} \in {\sf Pos}({\tt T})$.
\end{corollary}

\begin{proof}
By Proposition \ref{InsideRewriting} since ${\tt T} = {\tt T}[{\tt T}|_\mathbf{k}]_\mathbf{k}$ (Lemma \ref{LemmaSubtreeReplacement}).
\end{proof}

\subsection{Adequacy and completeness}

We aim to show that the tree rewriting calculi faithfully simulate logical derivations through the embeddings defined in Section \ref{sec:Embeddings}. Thereby, adequacy and completeness are key properties that underscore the efficacy of the tree rewriting calculi in relating logical inference and structural transformation.

\begin{theorem}[Completeness]
\label{Completeness}
Let ${\bf L}$ be a spi-logic such that ${\bf L} \subseteq \{({\sf 4}), ({\sf m}), ({\sf J})\}$. If $\varphi \vdash_{\bf L} \psi$, then ${\mathcal T}(\varphi) \hookrightarrow^{*}_{\sf{T}{\bf L}} {\mathcal T}(\psi)$ for $\varphi, \psi \in \mathscr{L}^{+}$.
\end{theorem}

\begin{proof}
By induction on the length of a ${\bf L}$ derivation. First, we handle the cases for $\bf K^+$-axioms and rules, and then conclude with those for ($\sf 4$), ($\sf m$), and ($\sf J$), by applying the corresponding rewriting rules straightforwardly.
    \begin{enumerate}[wide, labelwidth=!, labelindent=0pt]
    \item[] ${\mathcal T}(\varphi) \hookrightarrow^{*}_{\sf{T}{\bf L}} {\mathcal T}(\varphi)$ holds by no rewriting; ${\mathcal T}(\varphi) \hookrightarrow^{*}_{\sf{T}{\bf L}} {\mathcal T}(\top)$ by atom and child elimination.
    \item[] If ${\mathcal T}(\varphi) \hookrightarrow^{*}_{\sf{T}{\bf L}} {\mathcal T}(\psi)$ and ${\mathcal T}(\psi) \hookrightarrow^{*}_{\sf{T}{\bf L}} {\mathcal T}(\phi)$, then ${\mathcal T}(\varphi) \hookrightarrow^{*}_{\sf{T}{\bf L}} {\mathcal T}(\phi) \hookrightarrow^{*}_{\sf{T}{\bf L}} {\mathcal T}(\phi)$.
    \item[] ${\mathcal T}(\varphi \wedge \psi) = {\mathcal T} (\varphi) + {\mathcal T}(\psi) \hookrightarrow^{*}_{\sf{T}{\bf L}} {\mathcal T}(\varphi)$ and ${\mathcal T}(\varphi \wedge \psi) \hookrightarrow^{*}_{\sf{T}{\bf L}} {\mathcal T}(\psi)$ by Lemma \ref{LemmaRewSum}.
    \item[] If ${\mathcal T}(\varphi) \hookrightarrow^{*}_{\sf{T}{\bf L}} {\mathcal T}(\psi)$ and ${\mathcal T}(\varphi) \hookrightarrow^{*}_{\sf{T}{\bf L}} {\mathcal T}(\phi)$, by Lemma \ref{LemmaRewSum} we conclude ${\mathcal T}(\varphi) \hookrightarrow^{*}_{\sf{T}{\bf L}} {\mathcal T}(\psi) + {\mathcal T}(\phi) = {\mathcal T}(\psi \wedge \phi)$.
    \item[] If ${\mathcal T}(\varphi) \hookrightarrow^{*}_{\sf{T}{\bf L}} {\mathcal T}(\psi)$, then ${\mathcal T}(\langle \alpha \rangle \varphi) \hookrightarrow^{*}_{\sf{T}{\bf L}} {\mathcal T}(\langle \alpha \rangle \psi)$ by the Deep Rewriting Property (Corollary \ref{InsideRewritingRepl}).
    \item[($\sf 4$):] By the $\sf 4$-rule, ${\mathcal T}(\langle \alpha \rangle \langle \alpha \rangle \varphi) = \langle \varnothing ; [(\alpha,\langle \varnothing;[(\alpha,{\mathcal T}(\varphi))]\rangle)]\rangle \hookrightarrow^{\sf 4} \langle \varnothing;[(\alpha,{\mathcal T}(\varphi))] \rangle = {\mathcal T}(\langle \alpha \rangle \varphi)$.
    \item[($\sf m$):] Let $\alpha > \beta$. By the $\sf m$-rule, we show ${\mathcal T}(\langle \alpha \rangle \varphi) = \langle \varnothing ; [(\alpha, {\mathcal T}(\varphi))] \rangle \hookrightarrow^{\sf m} \langle \varnothing ; [(\beta, {\mathcal T}(\varphi))] \rangle = {\mathcal T}(\langle \beta \rangle \varphi)$.
    \item[($\sf J$):] Let $\alpha > \beta$ and ${\mathcal T}(\varphi) = \langle \Delta ; \Gamma \rangle$. By applying the $\sf J$-rule we see that
        \begin{equation*}
        \begin{split}
        {\mathcal T}(\langle \alpha \rangle \varphi \wedge \langle \beta \rangle \psi) & = \langle \varnothing;[(\alpha, \langle \Delta ; \Gamma \rangle),(\beta,{\mathcal T}(\psi))]\rangle \hookrightarrow^{\sf J} \langle \varnothing;[(\alpha,\langle \Delta ; \Gamma \smallfrown (\beta,{\mathcal T}(\psi)) \rangle)]\rangle \\
        & = \langle \varnothing;[(\alpha,{\mathcal T}(\varphi) + {\mathcal T}(\langle \beta \rangle \psi))] \rangle = {\mathcal T}(\langle \alpha \rangle (\varphi \wedge \langle \beta \rangle \psi)).
        \qedhere
        \end{split}    
        \end{equation*}
    \end{enumerate}
\end{proof}

To conclude, we show the adequacy of the tree rewriting calculi.

\begin{theorem}[Adequacy]
\label{Adequacy}
Let ${\bf L}$ be a spi-logic such that ${\bf L} \subseteq \{({\sf 4}), ({\sf m}), ({\sf J})\}$. If ${\tt T} \hookrightarrow^{*}_{{\sf T}{\bf L}} {\tt T'}$, then ${\mathcal F}({\tt T}) \vdash_{{\bf L}} {\mathcal F}({\tt T'})$ for ${\tt T}, {\tt T'} \in {\sf Tree^\diamond}$.
\end{theorem}

\begin{proof}
It suffices to show that if ${\tt T} \hookrightarrow_{{\sf T}{\bf L}} {\tt T'}$ then ${\mathcal F}({\tt T}) \vdash_{{\bf L}} {\mathcal F}({\tt T'})$, so we conclude by induction on the number of rewriting steps performed. Hence, we proceed by induction on ${\tt T}$. If ${\tt T}$ is a leaf, the result is trivially satisfied for the atomic rules, the only ones applicable. Otherwise, let ${\tt T} = \langle \Delta ; \Gamma \rangle$ for $\Gamma = [(\alpha_1,{\tt S}_1),...,(\alpha_m,{\tt S}_m)]$. Assuming ${\tt T} \hookrightarrow^{\mu} {\tt T'}$ for $\mu$ a rule of ${\sf T}{\bf L}$, we show ${\mathcal F}({\tt T}) \vdash_{\bf L} {\mathcal F}({\tt T'})$ by cases on the length of the position at which $\mu$ is applied. 

\medskip\noindent Suppose that $\mu$ is applied at $i\mathbf{k} \in {\sf Pos}({\tt T})$. Then, ${\tt T'}$ is of the form $\langle \Delta ; [(\alpha_1,{\tt S}_1),\\...,(\alpha_i,{\tt S'}),...,(\alpha_m,{\tt S}_m)] \rangle$ for ${\tt S}_i \hookrightarrow^{\mu} {\tt S'}$ at position $\mathbf{k}$. We conclude by the definition of $\mathcal F$, using the fact that ${\mathcal F}({\tt S}_i) \vdash_{{\bf L}} {\mathcal F}({\tt S'})$, which follows from the inductive hypothesis. The finer details are left to the reader.

\medskip\noindent Otherwise, if $\mu$ is applied at the empty position, we proceed by cases on $\mu$. Here, we provide detailed proofs for the 
$\sf 4$ and $\sf J$ rules, when present in ${\sf T}{\bf L}$, leaving the simpler cases to the reader.
\begin{enumerate}[wide, labelwidth=!, labelindent=0pt]
    \item[($\sf 4$-rule)]: Let ${\tt T} \hookrightarrow^{\sf 4} \langle \Delta ; \Gamma[(\beta, {\tt S})]_i \rangle$ for $\#_i \Gamma = (\beta, \langle \emptyset; [(\beta,{\tt S})] \rangle)$ and $0 < i \leq |\Gamma|$. By the axiom ($\sf 4$) of ${\bf L}$, we show that
    \begin{equation*}
    \begin{split}
    {\mathcal F}({\tt T}) & \equiv \bigwedge \Delta \wedge \langle \beta \rangle {\mathcal F}(\langle \emptyset; [(\beta,{\tt S})] \rangle) \wedge \bigwedge \Diamond (\Gamma^{-i}) \equiv \bigwedge \Delta \wedge \langle \beta \rangle \langle \beta \rangle {\mathcal F}({\tt S}) \wedge \bigwedge \Diamond (\Gamma^{-i}) \\
    & \vdash_{\bf L} \bigwedge \Delta \wedge \langle \beta \rangle {\mathcal F}({\tt S}) \wedge \bigwedge \Diamond (\Gamma^{-i}) \equiv {\mathcal F}(\langle \Delta ; \Gamma[(\beta, {\tt S})]_i \rangle).
    \end{split}
    \end{equation*}
    \item[($\sf J$-rule)]: Let ${\tt T} \hookrightarrow^{\sf J} \langle \Delta ; (\Gamma[(\alpha, \langle \tilde{\Delta} ; \tilde{\Gamma} \smallfrown (\beta , {\tt S}) \rangle)]_i)^{-j} \rangle$ for $\#_i \Gamma = (\alpha , \langle \tilde{\Delta} ; \tilde{\Gamma} \rangle)$, $\#_j \Gamma = (\beta, {\tt S})$ and $0 < i < j \leq |\Gamma|$ without loss of generality. By axiom ($\sf J$) of ${\bf L}$,
    \begin{equation*}
    \begin{split}
    {\mathcal F}({\tt T}) & \equiv \bigwedge \Delta \wedge \langle \alpha \rangle {\mathcal F}(\langle \tilde{\Delta} ; \tilde{\Gamma} \rangle) \wedge \langle \beta \rangle {\mathcal F}({\tt S}) \wedge \bigwedge \Diamond ((\Gamma^{-j})^{-i}) \\
    & \vdash_{\bf L} \bigwedge \Delta \wedge \langle \alpha \rangle ({\mathcal F}(\langle \tilde{\Delta} ; \tilde{\Gamma} \rangle) \wedge \langle \beta \rangle {\mathcal F}({\tt S})) \wedge \bigwedge \Diamond ((\Gamma^{-j})^{-i}) \\
    & \equiv \bigwedge \Delta \wedge \langle \alpha \rangle {\mathcal F}(\langle \tilde{\Delta} ; \tilde{\Gamma} \smallfrown (\beta,{\tt S}) \rangle) \wedge \bigwedge \Diamond ((\Gamma^{-j})^{-i}) \\
    & \equiv {\mathcal F}(\langle \Delta ; (\Gamma[(\alpha, \langle \tilde{\Delta} ; \tilde{\Gamma} \smallfrown (\beta , {\tt S})  \rangle)]_i)^{-j} \rangle).
    \qedhere
    \end{split}
    \end{equation*}
    \qedhere
\end{enumerate}
\end{proof}

As a corollary, logical derivability can be established by rewriting the trees in which the formulas are embedded into. Likewise, a rewriting can be shown by proving the corresponding sequence of the formulas that embed the concerned trees.

\begin{corollary}
\label{TRStoRC}
Let ${\bf L}$ be a spi-logic such that ${\bf L} \subseteq \{({\sf 4}), ({\sf m}), ({\sf J})\}$. If ${\mathcal T}(\varphi) \hookrightarrow^{*}_{{\sf T}{\bf L}} {\mathcal T}(\psi)$, then $\varphi \vdash_{\bf L} \psi$. Conversely, if ${\mathcal F}({\tt T}) \vdash_{\bf L} {\mathcal F}({\tt T'})$, then ${\tt T} \hookrightarrow^{*}_{{\sf T}{\bf L}} {\tt T'}$.
\end{corollary}

\begin{proof}
By the adequacy and completeness of ${\sf T}{\bf L}$ (Theorems \ref{Adequacy} and \ref{Completeness}) using the Embedding Composition Property (Proposition \ref{PropInvEmbedding}). 
\end{proof}

\section{Rewrite Normalization}
\label{sec: Normalization}

We provide a normalization procedure for the tree rewriting calculi, offering explicit upper bounds. This normalization enables the transformation of trees through a specific sequence of rules based on their kind. For spi-logics without axiom ($\sf J$), we ensure normalization at the cost of an exponential increase in rule applications. Otherwise, we obtain a double-exponential bound.

\begin{definition}[Normal rewriting]
Let ${\bf L}$ be a spi-logic such that ${\bf L} \subseteq \{({\sf 4}), ({\sf m}), ({\sf J})\}$. A list $\Omega$ of rules of ${\sf T}{\bf L}$ is a {\em normal rewriting sequence} if it is of the form
\begin{equation*}
    \Omega_{{+}} \smallfrown \Omega_{\diamond} \smallfrown \Omega_\delta \smallfrown \Omega_\rho \smallfrown \Omega_\sigma
\end{equation*} 
for $\Omega_{+}, \Omega_\diamond, \Omega_\delta, \Omega_\rho$ and $\Omega_\sigma$ lists of replicative, modal, decreasing, atomic and structural rules, respectively.
\end{definition}

The order of the kinds of rules in normal rewriting adheres to the following principles. First, it is not always the case that performing a rule before a replicative one can be equivalently reversed. Similarly, modal and decreasing rules cannot be interchanged (for example, it may be necessary to decrease the label of an edge in order to apply $\sf 4$). Finally, atomic and structural rules (for node labels and child permutation) are placed last since nodes and children may be removed in the previous steps.

\subsection{Tree metrics in rewriting}

We wish to provide a framework for bounding the rewriting process by analyzing the variations of tree metrics under rewriting. Recall that ${\sf w}({\tt T})$, ${\sf h}({\tt T})$ and ${\sf n}({\tt T})$ denote the width, height and number of nodes of ${\tt T} \in {\sf Tree^\diamond}$, respectively. Notably, the tree metrics decrease when applying decreasing rules. For the application of rules other than $\pi^+$ and $\sf J$, they remain unchanged.

\begin{observation}
\label{ObservationTreeMetrics}
If ${\tt T} \hookrightarrow^{\Omega} {\tt S}$ for $\Omega$ a list of rules from $\{{\sf m}, \pi^-, {\sf 4}, \rho^+, \rho^-, \sigma\}$, then ${\sf w}({\tt S}) \leq {\sf w}({\tt T})$, ${\sf h}({\tt S}) \leq {\sf h}({\tt T})$ and ${\sf n}({\tt S}) \leq {\sf n}({\tt T})$.
\end{observation}

When applying $\sf J$ and $\pi^+$, the tree metrics exhibit different behaviors. For $\sf J$ rewriting, the number of nodes remains invariant, while the height is linearly bounded by the number of rules. Conversely, for $\pi^+$ rewriting, the height remains unchanged, while the number of nodes is exponentially bounded. In both cases, the width is linearly bounded by the number of rules.

\begin{lemma}
\label{TreeMetrics}
Let $\Omega_+$ and $\Omega_{\sf J}$ be lists of replicative and $\sf J$ rules, respectively.

\medskip\noindent \textit{1.} If ${\tt T} \hookrightarrow^{\Omega_+} {\tt S}$, then ${\sf w}({\tt S}) \leq |\Omega_+| + {\sf w}({\tt T})$, ${\sf h}({\tt S}) = {\sf h}({\tt T})$ and ${\sf n}({\tt S}) \leq 2^{|\Omega_+|} \cdot {\sf n}({\tt T}) - (2^{|\Omega_+|} - 1)$.

\medskip\noindent \textit{2.} If ${\tt T} \hookrightarrow^{\Omega_{\sf J}} {\tt S}$, then ${\sf w}({\tt S}) \leq |\Omega_{\sf J}| + {\sf w}({\tt T})$, ${\sf h}({\tt S}) \leq |\Omega_{\sf J}| + {\sf h}({\tt T})$ and ${\sf n}({\tt S}) = {\sf n}({\tt T})$.
\end{lemma}

\begin{proofidea}
Each statement holds by induction on the length of the list. For the inductive step, we perform a second induction on the length of the position of the considered rule. To bound the number of nodes for the replicative rewriting, it suffices to see that if ${\tt T} \hookrightarrow^{\pi^+} {\tt S}$, then ${\sf n}({\tt S}) \leq 2 \cdot {\sf n}({\tt T}) - 1$.
\end{proofidea}

\begin{figure}[h]
\centering  
    \begin{subfigure}{0.2\textwidth}
            \begin{tikzpicture}  
            \node[state][label=right: \hspace{0.8cm}$\hookrightarrow^{\sf J}$] (b) at (0,0) {};
            \node[state] (c) [below left =of b] {};
            \node[state] (d) [below right =of b] {};
            \node[emptystate] (e) [below left =of c] {};
            \node[emptystate] (f) [below right =of c] {};
            \draw[-latex] (b) -- (c) node[fill=white,inner sep=2pt,midway] {{\small $\alpha$}};
            \draw[-latex] (b) -- (d) node[fill=white,inner sep=2pt,midway] {{\small $\beta$}};
            \draw[dashed,-] (c) edge[dashed] (e); 
            \draw[dashed,-] (c) edge[dashed] (f); 
            \draw[red,-latex] (e) -- (f)  node[fill=white,inner sep=2pt,midway, below, sloped] {\textcolor{red}{\small ${\sf w}({\tt T})$}};
            \end{tikzpicture}
    \end{subfigure}
    \begin{subfigure}{0.2\textwidth}
            \centering
            \begin{tikzpicture} 
            \node[state] (b) at (0,0) {};
            \node[state] (c) [below =of b] {};
            \node[state](d) [below right =of c] {};
            \node[emptystate] (e) [below left =of c] {};
            \node[emptystate] (f) [below =of c] {};
            \draw[-latex] (b) -- (c) node[fill=white,inner sep=2pt,midway] {{\small $\alpha$}};
            \draw[-latex] (c) -- (d) node[fill=white,inner sep=2pt,midway] {{\small $\beta$}};
            \draw[dashed,-] (c) edge[dashed] (e); 
            \draw[dashed,-] (c) edge[dashed] (f); 
            \draw[red,-latex] (e) -- (d)  node[fill=white,inner sep=2pt,midway, below,sloped] {\hspace{-0.1cm}\textcolor{red}{\small ${\sf w}({\tt T}) + 1$}};
            \end{tikzpicture}
    \end{subfigure} 
    \hspace{0.3cm}
    \begin{subfigure}{0.2\textwidth}
            \begin{tikzpicture}  
            \node[emptystate] (a) at (0,0) {};
            \node[emptystate] (e) [left =of a] {};
            \node[emptystate] (g) [below =of f] {};
            \node[state][label=right: \hspace{0.4cm}$\hookrightarrow^{\sf J}$] (b) [below =of a] {};
            \node[state] (c) [below left =of b] {};
            \node[state] (d) [below right =of b] {};
            \draw[-latex] (b) -- (c) node[fill=white,inner sep=2pt,midway] {{\small $\alpha$}};
            \draw[-latex] (b) -- (d) node[fill=white,inner sep=2pt,midway] {{\small $\beta$}};
            \draw[dashed,-] (a) edge[dashed] (b);
            \draw[red,-latex] (e) -- (c)  node[fill=white,inner sep=2pt,midway,left] {\textcolor{red}{\small ${\sf h}({\tt T})$}};
            \end{tikzpicture}
    \end{subfigure}
    \begin{subfigure}{0.2\textwidth}
            \centering
            \begin{tikzpicture} 
            \node[emptystate] (a) at (0,0) {};
            \node[emptystate] (e) [right =of a] {};
            \node[emptystate] (f) [below =of e] {};
            \node[emptystate] (g) [below =of f] {};
            \node[emptystate] (h) [below =of g] {};
            \node[state] (b) [below =of a] {};
            \node[state] (c) [below =of b] {};
            \node[state](d) [below =of c] {};
            \draw[-latex] (b) -- (c) node[fill=white,inner sep=2pt,midway] {{\small $\alpha$}};
            \draw[-latex] (c) -- (d) node[fill=white,inner sep=2pt,midway] {{\small $\beta$}};
            \draw[dashed,-] (a) edge[dashed] (b); 
            \draw[red,-latex] (e) -- (h)  node[fill=white,inner sep=2pt,midway] {\textcolor{red}{\small ${\sf h}({\tt T}) + 1$}};
            \end{tikzpicture}
\end{subfigure}
\caption{The $\sf J$-rule may increase the width and height of a tree.}
\label{fig:4Jrules}
\end{figure}
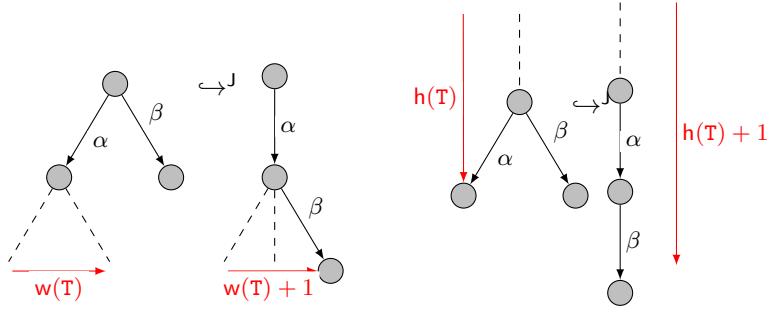

Conversely, the application of replicative rules is bounded by the increase in the number of nodes.

\begin{lemma}
\label{StarReplicativeNodes}
Let $\Omega_+$ be a list of replicative rules. If ${\tt T} \hookrightarrow^{\Omega_+} {\tt S}$, then $|\Omega_+| \leq {\sf n}({\tt S}) - {\sf n}({\tt T})$.
\end{lemma}

\begin{proofidea}
Since the number of nodes increases by at least one when applying $\pi^+$, we conclude by an easy induction on $|\Omega_+|$.
\end{proofidea}

\subsection{Rearrangement of replicative and structural rules}

It is desirable for the normalization process to reorder the application of replicative or structural rules, ensuring that they transform the tree in a desired sequence.

First, structural rules can be rearranged so that we first permute the children of deep nodes, and then the children of the root.

\begin{lemma}[Structural by depth]
\label{PermutationsFirstInsider}
Let $\Omega_\sigma$ be a list of structural rules. If ${\tt T} \hookrightarrow^{\Omega_\sigma} {\tt S}$, then we can rewrite ${\tt T} \hookrightarrow^{\Omega_\sigma^{> 0}} \circ \hookrightarrow^{\Omega_\sigma^{= 0}} {\tt S}$ for $\Omega_\sigma^{> 0}$ and $\Omega_\sigma^{= 0}$ lists of structural rules applied at non-empty and empty positions, respectively.
\end{lemma}

\begin{proofidea}
By induction on $|\Omega_\sigma|$. For the inductive step, we consider cases on the position of $\sigma$. To permute the application of $\sigma$ at the empty position with $\Omega_\sigma^{> 0}$, we swap the positions of the rules in the list that are applied within the subtrees permuted by $\sigma$.
\qedhere
\end{proofidea}

Using the previous rearrangement, we show that structural rewriting can be performed by a sequence that is linearly bounded by the number of nodes.

\begin{proposition}
\label{UBAnyPermutation}
If ${\tt T} \hookrightarrow^{\sigma *} {\tt S}$, then we can equivalently rewrite ${\tt T} \hookrightarrow^{\Omega_\sigma} {\tt S}$, where $\Omega_\sigma$ is a list of structural rules satisfying $|\Omega_\sigma| \leq {\sf n}({\tt T}) - 1$.
\end{proposition}

\begin{proof}
We proceed by induction on ${\tt T} \in {\sf Tree^\diamond}$, where the base case is trivially satisfied since no structural rule can be applied. Let ${\tt T} = \langle \Delta ; [(\alpha_1,{\tt S}_1),...,(\alpha_m,{\tt S}_m)]\rangle$. By Lemma \ref{PermutationsFirstInsider}, we have ${\tt T} \hookrightarrow^{\Omega^{> 0}_\sigma} \tilde{\tt S} \hookrightarrow^{\Omega_\sigma^{= 0}} {\tt S}$, where $\tilde{\tt S} = \langle \Delta ; [(\alpha_1,\tilde{\tt S}_1),...,(\alpha_m,\tilde{\tt S}_m)]\rangle$ and each subtree satisfies ${\tt S}_i \hookrightarrow^{\sigma *} \tilde{\tt S}_i$ for $1 \leq i \leq m$. By the inductive hypothesis, there exists a sequence $\tilde{\Omega}_\sigma^{i}$ of structural rules such that ${\tt S}_i \hookrightarrow^{\tilde{\Omega}_\sigma^{i}} \tilde{\tt S}_i$, satisfying $|\tilde{\Omega}_\sigma^i| \leq {\sf n}({\tt S}_i) - 1$. Applying the Deep Rewriting Property (Proposition \ref{InsideRewriting}), we obtain ${\tt T} \hookrightarrow^{\Omega_\sigma^1} \circ ... \circ \hookrightarrow^{\Omega_\sigma^m} \tilde{\tt S} \hookrightarrow^{\Omega_\sigma^{= 0}} {\tt S}$, where $\Omega_\sigma^i = [\sigma(i \mathbf{k},u,v) \mid \sigma(\mathbf{k},u,v) \in \tilde{\Omega}_\sigma^{i}]$ for $1 \leq i \leq m$. A well-known result in group theory \cite{lang2012algebra} states that any permutation of $m$ elements can be expressed as a product of at most $m-1$ transpositions. Since the application of $\Omega_\sigma^{= 0}$ corresponds to permuting the $m$ children of the root in $\tilde{\tt S}$, we can rewrite $\tilde{\tt S} \hookrightarrow^{\tilde{\Omega}_\sigma} {\tt S}$ for some sequence $\tilde{\Omega}_\sigma$ of structural rules applied at the root, with $|\tilde{\Omega}_\sigma| \leq m - 1$. Thus, the total number of applied rules satisfies $|\Omega_\sigma^1 \smallfrown ... \smallfrown \Omega_\sigma^m \smallfrown \tilde{\Omega}_\sigma| \leq \sum\limits_{i = 1}^{m} ({\sf n}({\tt S}_i) - 1) + m - 1 \leq {\sf n}({\tt T}) - 1$.
\end{proof}

To reorganize the application of replicative rules, we recall the concept of a \textit{sublist}, which preserves both the elements and their order within a given list. The following lemma establishes that, when applying replicative rules internally in the tree, we can choose to start or defer the application of the rules within a specific subtree.

\begin{lemma}
\label{LastDuplicateStarInternalLabel}
Let ${\tt T} \hookrightarrow^{\Omega_+} {\tt S}$ for $\Omega_+$ a list of replicative rules applied at non-empty positions, $\Omega_i$ the sublist of $\Omega_+$ consisting of rules applied at positions starting by $i \in {\sf Pos}({\tt T})$ and $\Omega_+ \setminus \Omega_i$ the sublist of $\Omega_+$ containing the remaining rules. We can equivalently rewrite ${\tt T} \hookrightarrow^{\Omega_i} \circ \hookrightarrow^{\Omega_+ \setminus \Omega_i} {\tt S}$ and ${\tt T} \hookrightarrow^{\Omega_+ \setminus \Omega_i} \circ \hookrightarrow^{\Omega_i} {\tt S}$.
\end{lemma}

\begin{proofidea}
Since the rules of $\Omega_+$ that are applied within the $i$-th child do not interact with the rest, we can straightforwardly permute their application.
\end{proofidea}

Rearranging replicative rules for progressively deeper applications within the tree results in a quadratic increase in the number of rules.

\begin{lemma}[Replicative by depth] \label{DuplicationsFirstMinDepth} Let ${\tt T} \hookrightarrow^{\Omega_+} {\tt S}$, where $\Omega_+$ is a list of replicative rules, and let $l \geq 0$ be the minimum length of the positions at which these rules are applied. Then, we have ${\tt T} \hookrightarrow^{\Omega_+^l} \circ \hookrightarrow^{\Omega_+^{> l}} {\tt S}$, where $\Omega_+^l$ is the list of replicative rules applied at positions of length $l$ and $\Omega_+^{> l}$ is the list of replicative rules applied at positions of length greater than $l$. Moreover, $|\Omega_+^l| \leq |\Omega_+|$ and $|\Omega_+^{> l}| \leq |\Omega_+|^2$. 
\end{lemma}

\begin{proofidea}
By induction on $|\Omega_+|$. For the inductive step, we proceed by induction on $l$. It suffices to show that if ${\tt T} \hookrightarrow^{\pi^+(\mathbf{k},i)} \circ \hookrightarrow^{\Omega_+^l} {\tt S}$ for $|\mathbf{k}| > l$, then ${\tt T} \hookrightarrow^{\Omega_+^l} \circ \hookrightarrow^{\Omega_+^{|\mathbf{k}|}} {\tt S}$, where $\Omega_+^{|\mathbf{k}|}$ is a list of replicative rules applied at positions of length $|\mathbf{k}|$ satisfying $|\Omega_+^{|\mathbf{k}|}| \leq 1 + |\Omega_+^l|$. The idea is similar to the one presented in Figure \ref{fig:Pi+Normalization}: the rules of $\Omega_+^{|\mathbf{k}|}$ are applied within the subtrees duplicated by $\Omega_+^l$ which were first transformed by the $\pi^+$-rule in the original derivation. We use Lemma \ref{LastDuplicateStarInternalLabel} to reorganize the rules in $\Omega_+^l$, allowing us to apply the inductive hypothesis and conclude by the Deep Rewriting Property (Proposition \ref{InsideRewriting}).
\end{proofidea}

\subsection{Permutation of rules}

We now turn to analyzing how different rewriting rules permute. Permuting rules generally requires intricate technical reasoning about tree positions and the arguments governing rule application. When two rules are applied at disjoint positions, their application permutes straightforwardly. However, when their applications overlap, additional rules may be required to resolve conflicts. To ensure clarity and insight into these scenarios, we include simplified examples alongside the proofs.

To begin, we explore how to permute a replicative rule with rules other than $\pi^+$ and $\sf J$. To reverse the application of a rule $\mu \in \mathscr{R} \setminus \{\pi^+, {\sf J}\}$ followed by a replicative rule, we may need an additional application of $\mu$ if $\mu$ was firstly applied within a subtree that was later duplicated by $\pi^+$ (see Figure \ref{fig:Pi+Normalization}). Recall that $\mathscr{R}$ denotes the set of all the rewriting rules of the tree rewriting calculi. As a result, we show that the application of a list of rules of $\mathscr{R} \setminus \{\pi^+, {\sf J}\}$ followed by a replicative rule can be equivalently reversed by doubling the list of rules.

\begin{lemma}
\label{PermutingReplicativeWithList}
Let $\Omega$ be a list of rules of $\mathscr{R} \setminus \{\pi^+, {\sf J}\}$. If ${\tt T} \hookrightarrow^{\Omega} \circ \hookrightarrow^{\pi^+} {\tt S}$, then we can rewrite it as ${\tt T} \hookrightarrow^{\pi^+} \circ \hookrightarrow^{\tilde{\Omega}} {\tt S}$, where $\tilde{\Omega}$ is a list of rules of $\mathscr{R} \setminus \{\pi^+, {\sf J}\}$ satisfying $|\tilde{\Omega}| \leq 2 \cdot |\Omega|$.
\end{lemma}

\begin{proof}
It suffices to establish that if ${\tt T} \hookrightarrow^{\mu} \circ \hookrightarrow^{\pi^+} {\tt S}$ for $\mu \in \mathscr{R} \setminus \{\pi^+, {\sf J}\}$, then either ${\tt T} \hookrightarrow^{\pi^+} \circ \hookrightarrow^{\mu} {\tt S}$ or ${\tt T} \hookrightarrow^{\pi^+} \circ \hookrightarrow^{\mu} \circ \hookrightarrow^{\mu} {\tt S}$. Once this is shown, we show the overall result by induction on $|\Omega|$. To prove this key result, we proceed by induction on the position of $\mu$, considering different cases based on the position of $\pi^+$. Here, we provide a detailed argument for the case where $\mu$ is $\rho^+$, leaving the remaining cases to the reader.

Let ${\tt T}$ be $\langle \Delta ; \Gamma \rangle$ for $\Gamma = [(\alpha_1,{\tt S}_1),...,(\alpha_m,{\tt S}_m)]$. We use $\hookrightarrow^{\rho^+(\mathbf{k},i)}$ to denote the application of $\rho^+$ at position $\mathbf{k}$, duplicating the $i$-th atom; and $\hookrightarrow^{\pi^+(\mathbf{l},j)}$ to denote the application of $\pi^+$ at position $\mathbf{l}$, duplicating the $j$-th child. Assuming ${\tt T} \hookrightarrow^{\rho^+(\mathbf{k},i)} \circ \hookrightarrow^{\pi^+(\mathbf{l},j)} {\tt S}$, we proceed by induction on $|\mathbf{k}|$. 
\begin{enumerate}[wide, labelwidth=!, labelindent=0pt]
    \item[\emph{Base case} ($\mathbf{k} = \epsilon$)]: We can easily rewrite it as ${\tt T} \hookrightarrow^{\pi^+(\mathbf{l},j)} \circ \hookrightarrow^{\rho^+(\epsilon,i)} {\tt S}$.
    \item[\emph{Inductive step} ($\mathbf{k} = r\mathbf{k'}$)]: The $\rho^+$-rule is applied within the subtree ${\tt S}_r$. We continue by cases on $|\mathbf{l}|$.
    \begin{enumerate}[wide, labelwidth=!, labelindent=10pt]
    \item[($\mathbf{l} = \epsilon$)]: If $j \neq r$, we simply reorder the application of the rules by modifying their arguments as ${\tt T} \hookrightarrow^{\pi^+(\epsilon,j)} \circ \hookrightarrow^{\rho^+((r + 1)\mathbf{k'},i)} {\tt S}$. Otherwise, we see that ${\tt T} \hookrightarrow^{\pi^+(\epsilon,r)} \circ \hookrightarrow^{\rho^+(1\mathbf{k'},i)} \circ \hookrightarrow^{\rho^+((r + 1)\mathbf{k'},i)} {\tt S}$.
    \item[($\mathbf{l} = n\mathbf{l'}$)]: If $r \neq n$, we trivially show ${\tt T} \hookrightarrow^{\pi^+(n\mathbf{l'},j)} \circ \hookrightarrow^{\rho^+(r\mathbf{k'},i)} {\tt S}$. Otherwise, given that ${\tt T} \hookrightarrow^{\rho^+(r\mathbf{k'},i)} \circ \hookrightarrow^{\pi^+(r\mathbf{l'},j)} {\tt S}$, we have ${\tt S}_r \hookrightarrow^{\rho^+(\mathbf{k'},i)} \circ \hookrightarrow^{\pi^+(\mathbf{l'},j)} {\tt S}|_r$. Then, by the inductive hypothesis we can equivalently rewrite it as ${\tt S}_r \hookrightarrow^{\pi^+} \circ \hookrightarrow^{\rho^+} {\tt S}|_r$ or ${\tt S}_r \hookrightarrow^{\pi^+} \circ \hookrightarrow^{\rho^+} \circ \hookrightarrow^{\rho^+} {\tt S}|_r$. Thus, we conclude that either ${\tt T} \hookrightarrow^{\pi^+} \circ \hookrightarrow^{\rho^+} {\tt S}$ or ${\tt T} \hookrightarrow^{\pi^+} \circ \hookrightarrow^{\rho^+} \circ \hookrightarrow^{\rho^+} {\tt S}$ by the Deep Rewriting Property (Proposition \ref{InsideRewriting}). 
    \qedhere
    \end{enumerate}
\end{enumerate}
\end{proof}

\begin{figure}[H]
    \centering
\begin{subfigure}{1\textwidth}
\centering
\begin{subfigure}{.4\textwidth}
\begin{subfigure}{.2\textwidth}
    \centering
    \begin{tikzpicture}   
    \node[state][label=right: \hspace{0.1cm}$\hookrightarrow^{\mu}$] (a) at (0,0) {} ;
    \node[emptystate] (c) [below =of a] {${\tt S}$};
    \draw[-latex] (a) -- (c) node[fill=white,inner sep=2pt,midway] {};
    \end{tikzpicture} 
\end{subfigure}
\hspace{0.2cm}
\begin{subfigure}{.1\textwidth}
    \centering
    \begin{tikzpicture}
    \node[state][label=right: \hspace{0.2cm}$\hookrightarrow^{\pi^+}$] (a) at (0,0) {} ;
    \node[emptystate] (c) [below =of a] {${\tt S}'$};
    \draw[-latex] (a) -- (c) node[fill=white,inner sep=2pt,midway] {};
    \end{tikzpicture} 
\end{subfigure}
\begin{subfigure}{.5\textwidth}
    \centering
    \begin{tikzpicture}        
    \node[state] (a) at (0,0) {} ;
    \node[emptystate] (b) [below left =of a] {${\tt S}'$};
    \node[emptystate] (c) [below right=of a] {${\tt S}'$};
    \draw[-latex] (a) -- (b) node[fill=white,inner sep=2pt,midway] {};
    \draw[-latex] (a) -- (c) node[fill=white,inner sep=2pt,midway] {};
    \end{tikzpicture} 
\end{subfigure}
\caption{${\tt T} \hookrightarrow^{\mu} \circ \hookrightarrow^{\pi^+} {\tt T}'$ for ${\tt S} \hookrightarrow^{\mu} {\tt S}'$.}
\end{subfigure}
\begin{subfigure}{.5\textwidth}
\centering
\begin{subfigure}{.1\textwidth}
    \centering
    \begin{tikzpicture}        
    \node[state][label=right: \hspace{0.2cm}$\hookrightarrow^{\pi^+}$] (a) at (0,0) {} ;
    \node[emptystate] (c) [below =of a] {${\tt S}$}; 
    \draw[-latex] (a) -- (c) node[fill=white,inner sep=2pt,midway] {};
    \end{tikzpicture} 
\end{subfigure}
\begin{subfigure}{.4\textwidth}
    \centering
    \begin{tikzpicture}        
    \node[state][label=right: \hspace{0.5cm}$\hookrightarrow^{\mu} \circ \hookrightarrow^{\mu} $] (a) at (0,0) {} ;
    \node[emptystate] (b) [below left =of a] {${\tt S}$};
    \node[emptystate] (c) [below right=of a] {${\tt S}$};
    \draw[-latex] (a) -- (b) node[fill=white,inner sep=2pt,midway] {};
    \draw[-latex] (a) -- (c) node[fill=white,inner sep=2pt,midway] {};
    \end{tikzpicture} 
\end{subfigure}
\begin{subfigure}{.4\textwidth}
    \centering
    \begin{tikzpicture}        
    \node[state] (a) at (0,0) {} ;
    \node[emptystate] (b) [below left =of a] {${\tt S}'$};
    \node[emptystate] (c) [below right=of a] {${\tt S}'$};
    \draw[-latex] (a) -- (b) node[fill=white,inner sep=2pt,midway] {};
    \draw[-latex] (a) -- (c) node[fill=white,inner sep=2pt,midway] {};
    \end{tikzpicture} 
\end{subfigure}
\caption{${\tt T} \hookrightarrow^{\pi^+} \circ \hookrightarrow^{\mu} \circ \hookrightarrow^{\mu} {\tt T}'$ for ${\tt S} \hookrightarrow^{\mu} {\tt S}'$.}
\end{subfigure}
\end{subfigure}
\caption{${\tt T} \hookrightarrow^{\mu} \circ \hookrightarrow^{\pi^+} {\tt T}'$ implies ${\tt T} \hookrightarrow^{\pi^+} \circ \hookrightarrow^{\mu} \circ \hookrightarrow^{\mu} {\tt T}'$ if $\mu \in \mathscr{R}$ is applied within the subtree that is duplicated.}
\label{fig:Pi+Normalization}
\end{figure}

\begin{figure}[h]
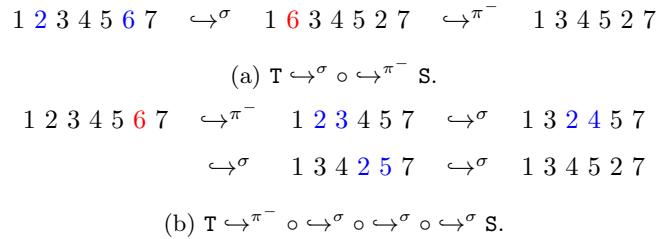

    \centering
    \begin{subfigure}{1.\textwidth} 
        \centering
        \begin{tabular}{c c c c c}
            1 \textcolor{blue}{2} 3 4 5 \textcolor{blue}{6} 7 & $\hookrightarrow^{\sigma}$ & 1 \textcolor{red}{6} 3 4 5 2 7 & $\hookrightarrow^{\pi^-}$ & 1 3 4 5 2 7
        \end{tabular}
        \caption{${\tt T} \hookrightarrow^{\sigma} \circ \hookrightarrow^{\pi^-} {\tt S}$.}
    \end{subfigure}
    \begin{subfigure}{1.\textwidth}
        \centering
        \begin{tabular}{c c c c c}
            1 2 3 4 5 \textcolor{red}{6} 7 & $\hookrightarrow^{\pi^-}$ & 1 \textcolor{blue}{2} \textcolor{blue}{3} 4 5 7 & $\hookrightarrow^{\sigma}$ & 1 3 \textcolor{blue}{2} \textcolor{blue}{4} 5 7 \\
            & $\hookrightarrow^{\sigma}$ & 1 3 4 \textcolor{blue}{2} \textcolor{blue}{5} 7 & $\hookrightarrow^{\sigma}$ & 1 3 4 5 2 7 
        \end{tabular}
        \caption{${\tt T} \hookrightarrow^{\pi^-} \circ \hookrightarrow^{\sigma} \circ \hookrightarrow^{\sigma} \circ \hookrightarrow^{\sigma} {\tt S}$.}
    \end{subfigure}
    \caption{Example of erasing child $6$ after being permuted with $2$.}
    \label{fig:ExampleStructuralPermutation}
\end{figure}

We proceed with the the study of structural rules. Notably, the application of $\sigma$ followed by a rule other than $\pi^+$, $\sf J$ or $\pi^-$ can be reversed straightforwardly. However, if one wants to permute the application of $\sigma$ followed by $\pi^-$ or $\sf J$, we require an additional sequence of structural rules that is linearly bounded by the width of the initial tree. This occurs when a subtree permuted by $\sigma$ is later erased using $\pi^-$ or $\sf J$. Note that this scenario also applies to the $\sf J$-rule, as it removes a child and relocates it to a deeper part of the tree. Figure \ref{fig:ExampleStructuralPermutation} illustrates an example of the sequence of $\sigma$-rules required in this cases.

By repeatedly permuting with $\sigma$ using the previous reasoning, we derive the following result for a sequence of structural rules.

\begin{lemma}
\label{PermuteWithStructural}
Let ${\tt T} \hookrightarrow^{\Omega_\sigma} \circ \hookrightarrow^{\mu} {\tt S}$ for $\Omega_\sigma$ a list of structural rules.

\medskip\noindent \textit{1.} If $\mu \in \{{\sf m}, {\sf 4}, \rho^+, \rho^-\}$, we can equivalently rewrite ${\tt T} \hookrightarrow^{\mu} \circ \hookrightarrow^{\Omega_\sigma} {\tt S}$.

\medskip\noindent \textit{2.} If $\mu \in \{\pi^-, {\sf J}\}$, we rewrite ${\tt T} \hookrightarrow^{\mu} \circ \hookrightarrow^{\tilde{\Omega}_\sigma} {\tt S}$, where $\tilde{\Omega}_\sigma$ is a list of structural rules such that $|\tilde{\Omega}_\sigma| \leq |\Omega_\sigma| \cdot {\sf w}({\tt T})$.
\end{lemma}

\begin{proof}
\begin{enumerate}[wide, labelwidth=!, labelindent=0pt]
\item By induction on $|\Omega_\sigma|$ using the fact that if ${\tt T} \hookrightarrow^{\sigma} \circ \hookrightarrow^{\mu} {\tt S}$, then ${\tt T} \hookrightarrow^{\mu} \circ \hookrightarrow^{\sigma} {\tt S}$. We further proceed by induction on the length of the position of $\sigma$, considering cases on the position of $\mu$; and leave the details to the reader.
    
\item It suffices to show that if ${\tt T} \hookrightarrow^{\sigma} \circ \hookrightarrow^{\mu} {\tt S}$, then ${\tt T} \hookrightarrow^{\mu} \circ \hookrightarrow^{\tilde{\Omega}_\sigma} {\tt S}$ such that $|\tilde{\Omega}_\sigma| \leq {\sf w}({\tt T})$, and conclude by induction on $|\Omega_\sigma|$. We show this key result for both cases of $\mu$.

\begin{enumerate}[wide, labelwidth=!, labelindent=0pt]
\item[($\mu = \pi^-$)]: We continue by induction on the length of the position of $\sigma$, considering cases on the position of $\pi^-$. Below, we outline the different scenarios of the case, leaving the details to the reader. 

If $\sigma$ is applied at a non-empty position, we see that ${\tt T} \hookrightarrow^{\pi^-} \circ \hookrightarrow^{\sigma} {\tt S}$ if the rules are applied disjointly, ${\tt T} \hookrightarrow^{\pi^-} {\tt S}$ if $\pi^-$ erases the child within $\sigma$ is applied, or ${\tt T} \hookrightarrow^{\pi^{-}} \circ \hookrightarrow^{\tilde{\Omega}_\sigma} {\tt S}$ satisfying $|\tilde{\Omega}_\sigma| \leq {\sf w}({\tt T})$ if both rules are applied within the same subtree using the inductive hypothesis and the Deep Rewriting Property (Proposition \ref{InsideRewriting}). Otherwise, if $\sigma$ is applied at the empty position, we show ${\tt T} \hookrightarrow^{\pi^-} \circ \hookrightarrow^{\sigma} {\tt S}$ except for the cases where $\pi^-$ erases a child which was firstly permuted by $\sigma$. Let us elaborate on these scenarios.

We write $\hookrightarrow^{\sigma(\epsilon,i,u)}$ to denote the application of $\sigma$ at position $\epsilon$ permuting the $i$-th and the $u$-th children, and $\hookrightarrow^{\pi^-(\epsilon,n)}$ to denote the application of $\pi^-$ at position $\epsilon$ erasing the $n$-th child. Without loss of generality, we assume ${\tt T} \hookrightarrow^{\sigma(\epsilon,i,u)} \circ \hookrightarrow^{\pi^-(\epsilon, i)} {\tt S}$ for $i < u$. If ${\tt T}$ has only two children, there is only one left in ${\tt S}$. Hence, since no reorganization is needed, we simply conclude ${\tt T} \hookrightarrow^{\pi^-(\epsilon, u)} {\tt S}$. Otherwise, we show 
\begin{equation*}
    {\tt T} \hookrightarrow^{\pi^{-}(\epsilon,u)} \circ \hookrightarrow^{\sigma(\epsilon,i,i+1)} \circ \hookrightarrow^{\sigma(\epsilon,i+1,i+2)} \circ ... \circ \hookrightarrow^{\sigma(\epsilon,u-2,u-1)} {\tt S}
\end{equation*} where $|\sigma(\epsilon,i,i+1) \smallfrown \sigma(\epsilon,i+1,i+2) \smallfrown ... \smallfrown \sigma(\epsilon,u-2,u-1)| = u - 2 - i \leq n - 2 \leq {\sf w}({\tt T})$.
\item[($\mu = {\sf J}$)]: As for the previous case, we proceed by induction on the length of the position of $\sigma$, considering cases on the position of $\sf J$. We outline the scenarios in a similar manner, leaving the details to the reader. If $\sigma$ is applied at a non-empty position, we either conclude ${\tt T} \hookrightarrow^{\sf J} \circ \hookrightarrow^{\sigma} {\tt S}$ if the rules are applied disjointly, or ${\tt T} \hookrightarrow^{\sf J} \circ \hookrightarrow^{\tilde{\Omega}_\sigma} {\tt S}$ such that $|\tilde{\Omega}_\sigma| \leq {\sf w}({\tt T})$ if both rules are applied within the same subtree. If $\sigma$ is applied at the empty position, we conclude ${\tt T} \hookrightarrow^{\sf J} \circ \hookrightarrow^{\sigma} {\tt S}$ except for the cases in which $\sigma$ permutes the child that is then relocated by $\sf J$. Thus, following the same reasoning as for $\pi^-$, we conclude that ${\tt T} \hookrightarrow^{\sf J} \circ \hookrightarrow^{\tilde{\Omega}_\sigma} {\tt S}$, where $|\tilde{\Omega}_\sigma| \leq {\sf w}({\tt T})$.
\qedhere
\end{enumerate}
\end{enumerate}
\end{proof}

Hence, we see that structural rules can always be delayed to the end of the rewriting.

\begin{corollary}
\label{CorollarySigma}
Let $\Omega$ be a list of rules of $\mathscr{R} \setminus \{\sigma\}$. If ${\tt T} \hookrightarrow^{\sigma *} \circ \hookrightarrow^{\Omega} {\tt S}$, then we can equivalently rewrite ${\tt T} \hookrightarrow^{\Omega} \circ \hookrightarrow^{\sigma *} {\tt S}$.
\end{corollary}

\begin{proof}
By induction on $|\Omega|$ using Lemmas \ref{PermutingReplicativeWithList} and  \ref{PermuteWithStructural} for the inductive step.
\end{proof}

We now examine how atomic rules interact with a decreasing or modal rule.

\begin{lemma}
\label{PermuteWithAtomic}
Let $\Omega_\rho$ be a list of atomic rules and $\mu \in \{\pi^-, {\sf 4}, {\sf m}, {\sf J}\}$. If ${\tt T} \hookrightarrow^{\Omega_\rho} \circ \hookrightarrow^{\mu} {\tt S}$, then we can rewrite ${\tt T} \hookrightarrow^{\mu} \circ \hookrightarrow^{\tilde{\Omega}_\rho} {\tt S}$, where $\tilde{\Omega}_\rho$ is a list of atomic rules satisfying $|\tilde{\Omega}_\rho| \leq |\Omega_\rho|$.
\end{lemma}

\begin{proof}
Let $\rho \in \{\rho^+, \rho^-\}$. Observe that if ${\tt T} \hookrightarrow^{\rho} \circ \hookrightarrow^{\mu} {\tt S}$ for $\mu \in \{{\sf 4}, {\sf m}, {\sf J}\}$, then ${\tt T} \hookrightarrow^{\mu} \circ \hookrightarrow^{\rho} {\tt S}$, since $\mu$ does not modify the labels of the nodes. Similarly, if ${\tt T} \hookrightarrow^{\rho} \circ \hookrightarrow^{\pi^-} {\tt S}$, then either ${\tt T} \hookrightarrow^{\pi^-} {\tt S}$ (if $\pi^-$ removes the node affected by $\rho$) or ${\tt T} \hookrightarrow^{\pi^-} \circ \hookrightarrow^{\rho} {\tt S}$. We prove both of these statements by induction on the length of the position of $\rho$, considering cases on the position of $\mu$. The finer details are left to the reader. Therefore, the result follows by an easy induction on $|\Omega_\rho|$.
\end{proof}

Reversing the application of a list of decreasing rules followed by a modal rule can cause the number of applied modal rules to grow exponentially.

\begin{lemma}
\label{PermuteWithDecreasing}
Let $\Omega_\delta$ be a list of decreasing rules and $\mu$ a modal rule. If ${\tt T} \hookrightarrow^{\Omega_\delta} \circ \hookrightarrow^{\mu} {\tt S}$, then we can rewrite ${\tt T} \hookrightarrow^{\Omega_\mu} \circ \hookrightarrow^{\Omega_\delta} {\tt S}$, where $\Omega_\mu$ is a list of $\mu$-rules such that $|\Omega_\mu| \leq 2^{|\Omega_\delta|}$.
\end{lemma}

\begin{proof}
We proceed by induction on $|\Omega_\delta|$, with the base case trivially satisfied. For the inductive step, it suffices to show that if ${\tt T} \hookrightarrow^{\delta} \circ \hookrightarrow^{\mu} {\tt S}$ for $\delta \in \{\pi^-, {\sf 4}\}$, then either ${\tt T} \hookrightarrow^{\mu} \circ \hookrightarrow^{\delta} {\tt S}$ or ${\tt T} \hookrightarrow^{\mu} \circ \hookrightarrow^{\mu} \circ \hookrightarrow^{\delta} {\tt S}$. The proof proceeds by induction on the length of the position of $\delta$, considering cases on the position of $\mu$. If $\delta$ is $\pi^-$, we simply see that ${\tt T} \hookrightarrow^{\mu} \circ \hookrightarrow^{\pi^-} {\tt S}$. Otherwise, if $\delta$ is $\sf 4$, we show that either ${\tt T} \hookrightarrow^{\mu} \circ \hookrightarrow^{\sf 4} {\tt S}$ or ${\tt T} \hookrightarrow^{\mu} \circ \hookrightarrow^{\mu} \circ \hookrightarrow^{\sf 4} {\tt S}$ for the scenarios displayed in Figures \ref{fig:4LambdaNormalization} and \ref{fig:4JNormalization}. The details are left to the reader.
\end{proof}

\begin{figure}[t]
    \centering
\begin{subfigure}{1\textwidth}
\begin{subfigure}{.4\textwidth}
\centering
\begin{subfigure}{.2\textwidth}
    \centering
    \begin{tikzpicture}   
     \node[state] (b) at (0,0) {};
     \node[state][label=right: \hspace{0.1cm}$\hookrightarrow^{\sf 4}$] (d) [below =of b] {};
     \node[state] (f) [below =of d] {};
     \draw[-latex] (b) -- (d) node[fill=white,inner sep=2pt,midway] {$\alpha$}; 
     \draw[-latex] (d) -- (f) node[fill=white,inner sep=2pt,midway] {$\alpha$};
    \end{tikzpicture} 
\end{subfigure}
\hspace{0.2cm}
\begin{subfigure}{.2\textwidth}
    \centering
    \begin{tikzpicture}   
    \node[state][label=right: \hspace{0.2cm}$\hookrightarrow^{\sf m}$] (b) at (0,0) {};
    \node[state] (d) [below =of b] {};
    \draw[-latex] (b) -- (d) node[fill=white,inner sep=2pt,midway] {$\alpha$}; 
    \end{tikzpicture} 
\end{subfigure}
\hspace{0.1cm}
\begin{subfigure}{.2\textwidth}
    \centering
    \begin{tikzpicture}   
    \node[state] (b) at (0,0) {};
    \node[state] (d) [below =of b] {};
    \draw[-latex] (b) -- (d) node[fill=white,inner sep=2pt,midway] {$\beta$}; 
    \end{tikzpicture} 
\end{subfigure}
\caption{${\tt T} \hookrightarrow^{\sf 4} \circ \hookrightarrow^{\sf m} {\tt S}$.}
\end{subfigure}
\begin{subfigure}{0.6\textwidth}
\centering
\begin{subfigure}{.15\textwidth}
    \centering
    \begin{tikzpicture}   
     \node[state] (b) at (0,0) {};
     \node[state][label=right: \hspace{0.1cm}$\hookrightarrow^{\sf m}$] (d) [below =of b] {};
     \node[state] (f) [below =of d] {};
     \draw[-latex] (b) -- (d) node[fill=white,inner sep=2pt,midway] {$\alpha$}; 
     \draw[-latex] (d) -- (f) node[fill=white,inner sep=2pt,midway] {$\alpha$};
    \end{tikzpicture} 
\end{subfigure}
\hspace{0.1cm}
\begin{subfigure}{.15\textwidth}
    \centering
    \begin{tikzpicture}   
     \node[state] (b) at (0,0) {};
     \node[state][label=right: \hspace{0.1cm}$\hookrightarrow^{\sf m}$] (d) [below =of b] {};
     \node[state] (f) [below =of d] {};
     \draw[-latex] (b) -- (d) node[fill=white,inner sep=2pt,midway] {$\beta$}; 
     \draw[-latex] (d) -- (f) node[fill=white,inner sep=2pt,midway] {$\alpha$};
    \end{tikzpicture} 
\end{subfigure}
\hspace{0.1cm}
\begin{subfigure}{.15\textwidth}
    \centering
    \begin{tikzpicture}   
     \node[state] (b) at (0,0) {};
     \node[state][label=right: \hspace{0.2cm}$\hookrightarrow^{\sf 4}$] (d) [below =of b] {};
     \node[state] (f) [below =of d] {};  
     \draw[-latex] (b) -- (d) node[fill=white,inner sep=2pt,midway] {$\beta$}; 
     \draw[-latex] (d) -- (f) node[fill=white,inner sep=2pt,midway] {$\beta$};
    \end{tikzpicture} 
\end{subfigure}
\hspace{0.1cm}
\begin{subfigure}{.08\textwidth}
    \centering
    \begin{tikzpicture}   
    \node[state] (b) at (0,0) {};
    \node[state] (d) [below =of b] {};
    \draw[-latex] (b) -- (d) node[fill=white,inner sep=2pt,midway] {$\beta$}; 
    \end{tikzpicture} 
\end{subfigure}
\caption{${\tt T} \hookrightarrow^{\sf m} \circ \hookrightarrow^{\sf m} \circ \hookrightarrow^{\sf 4} {\tt S}$.}
\end{subfigure}
\end{subfigure}
\caption{${\tt T} \hookrightarrow^{\sf 4} \circ \hookrightarrow^{\sf m} {\tt S}$ implies ${\tt T} \hookrightarrow^{\sf m} \circ \hookrightarrow^{\sf m} \circ \hookrightarrow^{\sf 4} {\tt S}$.}
\label{fig:4LambdaNormalization}
\end{figure}

\begin{figure}[t]
\centering
\begin{subfigure}{1\textwidth}
\centering
\begin{subfigure}{.2\textwidth}
    \centering
    \begin{tikzpicture}  
        \node[state] (b) at (0,0) {};
        \node[state] (c) [below left =of b] {};
        \node[state][label=right: \hspace{0.5cm}$\hookrightarrow^{\sf 4}$] (d) [below right =of b] {};
        \node[state] (f) [below =of c] {};
        \draw[-latex] (b) -- (c) node[fill=white,inner sep=2pt,midway] {{\small $\alpha$}};
        \draw[-latex] (c) -- (f) node[fill=white,inner sep=2pt,midway] {{\small $\alpha$}};
        \draw[-latex] (b) -- (d) node[fill=white,inner sep=2pt,midway] {{\small $\beta$}};
    \end{tikzpicture}
\end{subfigure}
\begin{subfigure}{.2\textwidth}
    \centering
    \begin{tikzpicture}  
        \node[state] (b) at (0,0) {};
        \node[state][label=right: \hspace{0.3cm}$\hookrightarrow^{\sf J}$] (d) [below right =of b] {};
        \node[state] (f) [below left =of b] {};
        \node[emptystate] (h) [below left=of f] {};
        \draw[-latex] (b) -- (f) node[fill=white,inner sep=2pt,midway] {{\small $\alpha$}};
        \draw[-latex] (b) -- (d) node[fill=white,inner sep=2pt,midway] {{\small $\beta$}}; 
    \end{tikzpicture} 
\end{subfigure}
\begin{subfigure}{.2\textwidth}
    \centering
    \begin{tikzpicture}   
     \node[state] (b) at (0,0) {};
     \node[state] (d) [below =of b] {};
     \node[state] (f) [below =of d] {};  
     \draw[-latex] (b) -- (d) node[fill=white,inner sep=2pt,midway] {$\alpha$}; 
     \draw[-latex] (d) -- (f) node[fill=white,inner sep=2pt,midway] {$\beta$};
    \end{tikzpicture} 
\end{subfigure}
\caption{${\tt T} \hookrightarrow^{\sf 4} \circ \hookrightarrow^{\sf J} {\tt S}$.}
\end{subfigure}
\begin{subfigure}{1\textwidth}
\centering
\begin{subfigure}{.2\textwidth}
    \centering
    \begin{tikzpicture}  
        \node[state] (b) at (0,0) {};
        \node[state] (c) [below left =of b] {};
        \node[state][label=right: \hspace{0.6cm}$\hookrightarrow^{\sf J}$] (d) [below right =of b] {};
        \node[state] (f) [below =of c] {};
        \draw[-latex] (b) -- (c) node[fill=white,inner sep=2pt,midway] {{\small $\alpha$}};
        \draw[-latex] (c) -- (f) node[fill=white,inner sep=2pt,midway] {{\small $\alpha$}};
        \draw[-latex] (b) -- (d) node[fill=white,inner sep=2pt,midway] {{\small $\beta$}};
    \end{tikzpicture} 
\end{subfigure}
\begin{subfigure}{.2\textwidth}
    \centering
    \begin{tikzpicture}   
        \node[state] (b) at (0,0) {};
        \node[state][label=right: \hspace{0.5cm}$\hookrightarrow^{\sf J}$] (c) [below =of b] {};
        \node[state] (d) [below left =of c] {};
        \node[state] (f) [below right =of c] {};
        \draw[-latex] (b) -- (c) node[fill=white,inner sep=2pt,midway] {{\small $\alpha$}};
        \draw[-latex] (c) -- (f) node[fill=white,inner sep=2pt,midway] {{\small $\beta$}};
        \draw[-latex] (c) -- (d) node[fill=white,inner sep=2pt,midway] {{\small $\alpha$}};
    \end{tikzpicture} 
\end{subfigure}
\begin{subfigure}{.1\textwidth}
    \centering
    \begin{tikzpicture}   
         \node[state] (b) at (0,0) {};
        \node[state]  (c) [below =of b] {};
        \node[state][label=right: \hspace{0.4cm}$\hookrightarrow^{\sf 4}$] (f) [below =of c] {};
        \node[state] (d) [below =of f] {};
        \draw[-latex] (b) -- (c) node[fill=white,inner sep=2pt,midway] {{\small $\alpha$}};
        \draw[-latex] (c) -- (f) node[fill=white,inner sep=2pt,midway] {{\small $\alpha$}};
        \draw[-latex] (f) -- (d) node[fill=white,inner sep=2pt,midway] {{\small $\beta$}};
    \end{tikzpicture} 
\end{subfigure}
\begin{subfigure}{.1\textwidth}
    \centering
    \begin{tikzpicture}   
    \node[state] (b) at (0,0) {};
     \node[state] (d) [below =of b] {};
     \node[state] (f) [below =of d] {};  
     \draw[-latex] (b) -- (d) node[fill=white,inner sep=2pt,midway] {$\alpha$}; 
     \draw[-latex] (d) -- (f) node[fill=white,inner sep=2pt,midway] {$\beta$};
    \end{tikzpicture} 
\end{subfigure}
\caption{${\tt T} \hookrightarrow^{\sf J} \circ \hookrightarrow^{\sf J} \circ \hookrightarrow^{\sf 4} {\tt S}$.}
\end{subfigure}
\caption{${\tt T} \hookrightarrow^{\sf 4} \circ \hookrightarrow^{\sf J} {\tt S}$ implies ${\tt T} \hookrightarrow^{\sf J} \circ \hookrightarrow^{\sf J} \circ \hookrightarrow^{\sf 4} {\tt S}$.}
\label{fig:4JNormalization}
\end{figure}

\subsection{Interaction of modal and replicative rules}

Next, we examine how modal and replicative rules interact, beginning with $\sf J$ and $\pi^+$. When the applications of these two rules overlap, more complex situations can arise, where the application of both rules increases. To manage the interaction of multiple applications of these rules, we introduce the concept of \textit{flags} for specific nodes in the tree. These flags serve as markers placed on certain nodes in ${\tt T}$ when ${\tt T} \hookrightarrow^{\sf J} {\tt S}$. The purpose of the flags is to track which nodes are directly affected by a $\sf J$-transformation through the rewriting process. With the flags in place, we can effectively handle cases where the applications of $\pi^+$ and $\sf J$ overlap.

\begin{definition}[$\sf J$-flags]
\label{Jflags}
The {\em internal}, {\em upper}, and {\em lower $\sf J$-flags} of ${\tt T} \hookrightarrow^{\sf J} {\tt S}$ are placed in ${\tt T}$ at positions defined inductively on the length of the position of $\sf J$.
\begin{enumerate}[wide, labelwidth=!, labelindent=0pt]
\item[] Let $\sf J$ be applied at $\epsilon \in {\sf Pos}({\tt T})$ to the $i$-th and $j$-th children, with edges labelled $\alpha$ and $\beta$, respectively, and $\alpha > \beta$. The internal $\sf J$-flag is placed at $\epsilon$, the upper $\sf J$-flag at $i \in {\sf Pos}({\tt T})$, and the lower $\sf J$-flag at $j \in {\sf Pos}({\tt T})$.
\item[] Let $\sf J$ be applied at $l \mathbf{k} \in {\sf Pos}({\tt T})$. The internal $\sf J$-flag is placed at $l \in {\sf Pos}({\tt T})$, the upper $\sf J$-flag at $l\mathbf{n} \in {\sf Pos}({\tt T})$, and the lower $\sf J$-flag at $l\mathbf{m} \in {\sf Pos}({\tt T})$, where $\mathbf{n} \in {\sf Pos}({\tt T}|_l)$ and $\mathbf{m} \in {\sf Pos}({\tt T}|_l)$ are the positions of the upper and lower $\sf J$-flags of ${\tt T}|_l \hookrightarrow^{\sf J} {\tt S}|_l$, respectively.
\end{enumerate}
\end{definition}

\begin{figure}[t]
\centering  
    \begin{subfigure}{0.5\textwidth}
    \centering
        \begin{subfigure}{0.3\textwidth}
            \begin{tikzpicture} 
            \node[state][label=right: {\small $\epsilon$}] (x) at (0,0) {};
            \node[state][fill = black] (a) [below =of x] {};
            \node[state][label=right: \hspace{1cm} $\hookrightarrow^{\sf J}$] (b) [below =of a] {};
            \node[state][pattern = vertical lines] (c) [below left =of b] {};
            \node[state][pattern = horizontal lines] (d) [below right =of b] {};
            \draw[-latex] (b) -- (c) node[fill=white,inner sep=2pt,midway] {{\small $\alpha$}};
            \draw[-latex] (b) -- (d) node[fill=white,inner sep=2pt,midway] {{\small $\beta$}};
            \draw[dashed,-] (a) edge[dashed] (b);
            \draw[-latex] (x) -- (a) node[fill=white,inner sep=2pt,midway] {};
            \end{tikzpicture}
        \end{subfigure}
        \hspace{0.6cm}
        \begin{subfigure}{0.3\textwidth}
            \centering
            \begin{tikzpicture} 
            \node[state][label=right: {\small $\epsilon$}] (x) at (0,0) {};
            \node[state][fill = black] (a) [below =of x] {};
            \node[state] (b) [below =of a] {};
            \node[state][pattern = vertical lines] (c) [below =of b] {};
            \node[state][pattern = horizontal lines](d) [below =of c] {};
            \draw[-latex] (b) -- (c) node[fill=white,inner sep=2pt,midway] {{\small $\alpha$}};
            \draw[-latex] (c) -- (d) node[fill=white,inner sep=2pt,midway] {{\small $\beta$}};
            \draw[dashed,-] (a) edge[dashed] (b);
            \draw[-latex] (x) -- (a) node[fill=white,inner sep=2pt,midway] {};
            \end{tikzpicture}
        \end{subfigure} 
    \end{subfigure}
\caption{Internal $\sf J$-flag (black), upper $\sf J$-flag (vertical lines) and lower $\sf J$-flag (horizontal lines).}
\label{fig:JflagsEx}
\end{figure}
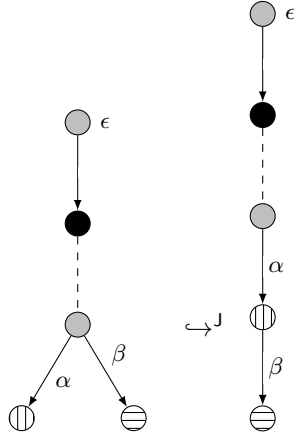

We now take a moment to describe how the $\sf J$-flags propagate through the rewriting process as markers. Our approach relies on the idea that placing a flag at a position can be interpreted as adding a fresh propositional variable to the list that labels the node at that position. As a result, during rewriting, the flag moves through the tree along with the node to which it was originally attached according to Definition \ref{Jflags}. Specifically, Figure \ref{fig:JflagsEx} illustrates how $\sf J$-flags move within the tree after the $\sf J$-rule is applied. Notably, when applying replicative rules, the occurrences of the flags may increase as the corresponding nodes are duplicated. We say that \textit{a flag is duplicated} when the node at the flag's position is duplicated. In contrast, when structural or modal rules are applied, the number of flags remains constant, though their positions in the tree change according to the transformation performed.

We observe how $\pi^+$ and $\sf J$ interact depending on the occurrences of the $\sf J$-flags.

\begin{proposition}
\label{ChildDupModal}
If ${\tt T} \hookrightarrow^{\sf J} \circ \hookrightarrow^{\pi^+} {\tt S}$, we can rewrite ${\tt T} \hookrightarrow^{\pi^+ *} \circ \hookrightarrow^{{\sf J} *} \circ \hookrightarrow^{\sigma *} {\tt S}$.
\end{proposition}

\begin{proof}
If $\pi^+$ does not increase the occurrences of the $\sf J$-flags, the transformations of the rules are disjoint, and we can directly rewrite ${\tt T} \hookrightarrow^{\pi^+} \circ \hookrightarrow^{\sf J} {\tt S}$. Otherwise, the rules overlap. Then, we can rewrite it as ${\tt T} \hookrightarrow^{\pi^+} \circ \hookrightarrow^{\sf J} \circ \hookrightarrow^{\sf J} {\tt S}$ when $\pi^+$ duplicates a subtree within which $\sf J$ was applied (see Figure \ref{fig:Pi+Normalization}), ${\tt T} \hookrightarrow^{\pi^+} \circ \hookrightarrow^{\pi^+} \circ \hookrightarrow^{\sf J} \circ \hookrightarrow^{\sf J} {\tt S}$ when $\pi^+$ duplicates the upper $\sf J$-flag (see Figure \ref{fig:JPi+1Normalization}), or ${\tt T} \hookrightarrow^{\pi^+} \circ \hookrightarrow^{\sf J} \circ \hookrightarrow^{\sf J} \circ \hookrightarrow^{\sigma *} {\tt S}$ when $\pi^+$ only duplicates the lower $\sf J$-flag (see Figure \ref{fig:JPi+2Normalization}). The proof proceeds by induction on the length of the position of $\sf J$, considering cases on the position of $\pi^+$. We leave the details to the reader.
\end{proof}

\begin{figure}[t]
    \centering
\begin{subfigure}{1\textwidth}
\centering
\begin{subfigure}{.3\textwidth}
    \centering
    \begin{tikzpicture}        
        \node[state][label=right: \hspace{0.9cm}$\hookrightarrow^{\sf J}$]  (b) at (0,0) {};
        \node[state] (c) [below left =of b] {};
        \node[state](d) [below right =of b] {};
        \draw[-latex] (b) -- (c) node[fill=white,inner sep=2pt,midway] {{\small $\alpha$}};
        \draw[-latex] (b) -- (d) node[fill=white,inner sep=2pt,midway] {{\small $\beta$}}; 
    \end{tikzpicture} 
\end{subfigure}
\begin{subfigure}{.1\textwidth}
    \centering
    \begin{tikzpicture}
    \node[state] (b) at (0,0) {};
     \node[state][label=right: \hspace{0.5cm}$\hookrightarrow^{\pi^+}$] (d) [below =of b] {};
     \node[state] (f) [below =of d] {}; 
     \draw[-latex] (b) -- (d) node[fill=white,inner sep=2pt,midway] {$\alpha$}; 
     \draw[-latex] (d) -- (f) node[fill=white,inner sep=2pt,midway] {$\beta$};
    \end{tikzpicture} 
\end{subfigure}
\hspace{0.2cm}
\begin{subfigure}{.2\textwidth}
    \centering
    \begin{tikzpicture}        
    \node[state] (b) at (0,0) {};
     \node[state] (d) [below left =of b] {};
     \node[state] (x) [below right =of b] {};
     \node[state] (f) [below =of d] {}; 
     \node[state] (z) [below =of x] {};
     \draw[-latex] (b) -- (d) node[fill=white,inner sep=2pt,midway] {$\alpha$}; 
     \draw[-latex] (d) -- (f) node[fill=white,inner sep=2pt,midway] {$\beta$};
     \draw[-latex] (b) -- (x) node[fill=white,inner sep=2pt,midway] {$\alpha$}; 
     \draw[-latex] (x) -- (z) node[fill=white,inner sep=2pt,midway] {$\beta$};
    \end{tikzpicture} 
\end{subfigure}
\caption{${\tt T} \hookrightarrow^{\sf J} \circ \hookrightarrow^{\pi^+} {\tt S}$.}
\end{subfigure}
\begin{subfigure}{1\textwidth}
\centering
\begin{subfigure}{.2\textwidth}
    \centering
    \begin{tikzpicture}        
        \node[state][label=right: \hspace{0.8cm}$\hookrightarrow^{\pi^+} \circ \hookrightarrow^{\pi^+}$] (b) at (0,0) {};
        \node[state] (c) [below left =of b] {};
        \node[state] (d) [below right =of b] {};
        \draw[-latex] (b) -- (c) node[fill=white,inner sep=2pt,midway] {{\small $\alpha$}};
        \draw[-latex] (b) -- (d) node[fill=white,inner sep=2pt,midway] {{\small $\beta$}}; 
    \end{tikzpicture} 
\end{subfigure}
\hspace{0.2cm}
\begin{subfigure}{.2\textwidth}
    \centering
    \begin{tikzpicture}        
        \node[state][label=right: \hspace{0.6cm}$\hookrightarrow^{\sf J}$] (b) at (0,0) {};
        \node[state] (c) [below =of b] {};
        \node[state] (d) [below right =of b] {};
        \node[state] (x) [below left =of b] {};
        \node[state] (y) [left =of x] {};
        \draw[-latex] (b) -- (c) node[fill=white,inner sep=2pt,midway] {{\small $\alpha$}};
        \draw[-latex] (b) -- (y) node[fill=white,inner sep=2pt,midway] {{\small $\alpha$}};
        \draw[-latex] (b) -- (d) node[fill=white,inner sep=2pt,midway] {{\small $\beta$}}; 
        \draw[-latex] (b) -- (x) node[fill=white,inner sep=2pt,midway] {{\small $\beta$}};
    \end{tikzpicture} 
\end{subfigure}
\hspace{1cm}
\begin{subfigure}{.2\textwidth}
    \centering
    \begin{tikzpicture}        
    \node[state] (b) at (0,0) {};
     \node[state] (d) [below left =of b] {};
     \node[state] (x) [below =of b] {};
     \node[state] (f) [below =of d] {}; 
     \node[state][label=right: \hspace{0.6cm}$\hookrightarrow^{\sf J}$] (z) [below right =of b] {}; 
     \draw[-latex] (b) -- (d) node[fill=white,inner sep=2pt,midway] {$\alpha$}; 
     \draw[-latex] (d) -- (f) node[fill=white,inner sep=2pt,midway] {$\beta$};
     \draw[-latex] (b) -- (x) node[fill=white,inner sep=2pt,midway] {$\alpha$}; 
     \draw[-latex] (b) -- (z) node[fill=white,inner sep=2pt,midway] {$\beta$};
    \end{tikzpicture} 
\end{subfigure}
\hspace{0.5cm}
\begin{subfigure}{.2\textwidth}
    \centering
    \begin{tikzpicture}        
    \node[state] (b) at (0,0) {};
     \node[state] (d) [below left =of b] {};
     \node[state] (x) [below right =of b] {};
     \node[state] (f) [below =of d] {}; 
     \node[state] (z) [below =of x] {}; 
     \draw[-latex] (b) -- (d) node[fill=white,inner sep=2pt,midway] {$\alpha$}; 
     \draw[-latex] (d) -- (f) node[fill=white,inner sep=2pt,midway] {$\beta$};
     \draw[-latex] (b) -- (x) node[fill=white,inner sep=2pt,midway] {$\alpha$}; 
     \draw[-latex] (x) -- (z) node[fill=white,inner sep=2pt,midway] {$\beta$};
    \end{tikzpicture} 
\end{subfigure}
\caption{${\tt T} \hookrightarrow^{\pi^+} \circ \hookrightarrow^{\pi^+} \circ \hookrightarrow^{\sf J} \circ \hookrightarrow^{\sf J} {\tt S}$.}
\end{subfigure}
\caption{${\tt T} \hookrightarrow^{\sf J} \circ \hookrightarrow^{\pi^+} {\tt S}$ implies ${\tt T} \hookrightarrow^{\pi^+} \circ \hookrightarrow^{\pi^+} \circ\hookrightarrow^{\sf J} \circ \hookrightarrow^{\sf J} {\tt S}$.}
\label{fig:JPi+1Normalization}
\end{figure}

\begin{figure}[t]
    \centering
\begin{subfigure}{1\textwidth}
\centering
\begin{subfigure}{.2\textwidth}
    \centering
    \begin{tikzpicture}  
        \node[state] (b) at (0,0) {};
        \node[state] (c) [below left =of b] {};
        \node[state][label=right: \hspace{0.4cm}$\hookrightarrow^{\sf J}$] (d) [below right =of b] {};
        \node[state] (e) [below =of c] {};
        \draw[-latex] (b) -- (c) node[fill=white,inner sep=2pt,midway] {{\small $\alpha$}};
        \draw[-latex] (b) -- (d) node[fill=white,inner sep=2pt,midway] {{\small $\beta$}};
        \draw[-latex] (c) -- (e) node[fill=white,inner sep=2pt,midway] {{\small $\gamma$}}; 
    \end{tikzpicture}
\end{subfigure}
\begin{subfigure}{.2\textwidth}
    \centering
    \begin{tikzpicture}   
     \node[state] (b) at (0,0) {};
     \node[state][label=right: \hspace{1.1cm}$\hookrightarrow^{\pi^+}$] (d) [below =of b] {};
     \node[state] (f) [below left =of d] {}; 
     \node[state] (g) [below right =of d] {};
     \draw[-latex] (b) -- (d) node[fill=white,inner sep=2pt,midway] {$\alpha$}; 
     \draw[-latex] (d) -- (f) node[fill=white,inner sep=2pt,midway] {$\gamma$}; 
     \draw[-latex] (d) -- (g) node[fill=white,inner sep=2pt,midway] {$\beta$};
    \end{tikzpicture} 
\end{subfigure}
\begin{subfigure}{.2\textwidth}
    \centering
    \begin{tikzpicture}   
     \node[state] (b) at (0,0) {};
     \node[state] (d) [below =of b] {};
     \node[state] (f) [below =of d] {}; 
     \node[state] (g) [below right =of d] {};
     \node[state] (h) [below left =of d] {};
     \draw[-latex] (b) -- (d) node[fill=white,inner sep=2pt,midway] {$\alpha$}; 
     \draw[-latex] (d) -- (f) node[fill=white,inner sep=2pt,midway] {$\gamma$}; 
     \draw[-latex] (d) -- (g) node[fill=white,inner sep=2pt,midway] {$\beta$};
     \draw[-latex] (d) -- (h) node[fill=white,inner sep=2pt,midway] {$\beta$};
    \end{tikzpicture} 
\end{subfigure}
\caption{${\tt T} \hookrightarrow^{\sf J} \circ \hookrightarrow^{\pi^+} {\tt S}$.}
\end{subfigure}
\begin{subfigure}{1.\textwidth}
\centering
\begin{subfigure}{.2\textwidth}
    \begin{tikzpicture}  
        \node[state] (b) at (0,0) {};
        \node[state] (c) [below left =of b] {};
        \node[state][label=right: \hspace{0.5cm}$\hookrightarrow^{\pi^+}$] (d) [below right =of b] {};
        \node[state] (e) [below =of c] {};
        \draw[-latex] (b) -- (c) node[fill=white,inner sep=2pt,midway] {{\small $\alpha$}};
        \draw[-latex] (b) -- (d) node[fill=white,inner sep=2pt,midway] {{\small $\beta$}};
        \draw[-latex] (c) -- (e) node[fill=white,inner sep=2pt,midway] {{\small $\gamma$}};
    \end{tikzpicture} 
\end{subfigure}
\hspace{0.8cm}
\begin{subfigure}{.2\textwidth}
    \centering
    \begin{tikzpicture}   
        \node[state] (b) at (0,0) {};
        \node[state] (c) [below =of b] {};
        \node[state][label=right: \hspace{0.4cm}$\hookrightarrow^{\sf J} \circ \hookrightarrow^{\sf J} $] (d) [below right =of b] {};
        \node[state] (e) [below left =of b] {};
        \node[state] (f) [below =of c] {};
        \draw[-latex] (b) -- (c) node[fill=white,inner sep=2pt,midway] {{\small $\alpha$}};
        \draw[-latex] (b) -- (d) node[fill=white,inner sep=2pt,midway] {{\small $\beta$}};
        \draw[-latex] (b) -- (e) node[fill=white,inner sep=2pt,midway] {{\small $\beta$}};
        \draw[-latex] (c) -- (f) node[fill=white,inner sep=2pt,midway] {{\small $\gamma$}};
    \end{tikzpicture} 
\end{subfigure}
\hspace{0.6cm}
\begin{subfigure}{.1\textwidth}
    \centering
    \begin{tikzpicture}   
    \node[state] (b) at (0,0) {};
     \node[state][label=right: \hspace{1cm}$\hookrightarrow^{\sigma}$] (d) [below =of b] {};
     \node[state] (f) [below =of d] {}; 
     \node[state] (g) [below right =of d] {};
     \node[state] (h) [below left =of d] {};
     \draw[-latex] (b) -- (d) node[fill=white,inner sep=2pt,midway] {$\alpha$}; 
     \draw[-latex] (d) -- (f) node[fill=white,inner sep=2pt,midway] {$\beta$}; 
     \draw[-latex] (d) -- (g) node[fill=white,inner sep=2pt,midway] {$\beta$};
     \draw[-latex] (d) -- (h) node[fill=white,inner sep=2pt,midway] {$\gamma$};
    \end{tikzpicture} 
\end{subfigure}
\hspace{1.5cm}
\begin{subfigure}{.1\textwidth}
    \centering
    \begin{tikzpicture}   
     \node[state] (b) at (0,0) {};
     \node[state] (d) [below =of b] {};
     \node[state] (f) [below =of d] {}; 
     \node[state] (g) [below right =of d] {};
     \node[state] (h) [below left =of d] {};
     \draw[-latex] (b) -- (d) node[fill=white,inner sep=2pt,midway] {$\alpha$}; 
     \draw[-latex] (d) -- (f) node[fill=white,inner sep=2pt,midway] {$\gamma$}; 
     \draw[-latex] (d) -- (g) node[fill=white,inner sep=2pt,midway] {$\beta$};
     \draw[-latex] (d) -- (h) node[fill=white,inner sep=2pt,midway] {$\beta$};
    \end{tikzpicture} 
\end{subfigure}
\caption{${\tt T} \hookrightarrow^{\pi^+} \circ \hookrightarrow^{\sf J} \circ \hookrightarrow^{\sf J} \circ \hookrightarrow^{\sigma} {\tt S}$.}
\end{subfigure}
\caption{${\tt T} \hookrightarrow^{\sf J} \circ \hookrightarrow^{\pi^+} {\tt S}$ implies ${\tt T} \hookrightarrow^{\pi^+} \circ \hookrightarrow^{\sf J} \circ \hookrightarrow^{\sf J} \circ \hookrightarrow^{\sigma} {\tt S}$.}
\label{fig:JPi+2Normalization}
\end{figure}

Permuting a list of modal rules with a replicative rule is particularly challenging because both the number of $\sf J$ and $\pi^+$ applications can increase when they permute. As a result, a straightforward induction on the number of modal rules does not terminate. To overcome this issue, we first show how to permute a single $\sf J$ with a list of replicative rules, ensuring termination by using induction on the occurrences of $\sf J$-flags. To achieve this, we need to reorganize the application of replicative rules.

Additionally, we need to rearrange replicative rules to start by duplicating a flag. In particular, we rearrange rules applied at a fixed depth level of the tree. To accomplish this reorganization, we may need to apply structural rules to ensure the proper order of the children.

\begin{lemma}
\label{DuplicationsAtWidth}
Let ${\tt T} \hookrightarrow^{\Omega_+} {\tt S}$ for $\Omega_+$ a list of replicative rules applied at positions of the same length that duplicates a flag of ${\tt T}$. Then, we can equivalently rewrite it as ${\tt T} \hookrightarrow^{\pi^+} \circ \hookrightarrow^{\hat{\Omega}_+} \circ \hookrightarrow^{\sigma *} {\tt S}$. Here, the rules are applied at positions of the same length as those of $\Omega_+$, the flag is duplicated in the first $\pi^+$-rule and $\hat{\Omega}_+$ is a list of replicative rules satisfying $|\hat{\Omega}_+| = |\Omega_+| - 1$.
\end{lemma}

\begin{proofidea}
By induction on the length of the positions in $\Omega_+$. For the inductive step, we reorganize $\Omega_+$ using Lemma \ref{LastDuplicateStarInternalLabel} to start applying replicative rules within the subtree containing the flag. We then apply the inductive hypothesis to this subtree and conclude using the Deep Rewriting Property (Proposition \ref{InsideRewriting}). Since, by definition, $\pi^+$ places the duplicated child at the beginning of the list of children, we may need to apply $\sigma$ to achieve the order of the original derivation.
However, this issue is solved by Corollary \ref{CorollarySigma}, which allows us to move the structural rules to the end of the rewriting.
\end{proofidea}

Building on the reorganization of replicative rules, one can choose to start duplicating a $\sf J$-flag after the $\sf J$-rule.

\begin{proposition} 
\label{StartDupJFlag} 
Let ${\tt T} \hookrightarrow^{\sf J} \circ \hookrightarrow^{\Omega_+} {\tt S}$, where either the upper $\sf J$-flag or the lower $\sf J$-flag occurs more than once in ${\tt S}$. Then, ${\tt T} \hookrightarrow^{\pi^+ *} \circ \hookrightarrow^{\sf J} \circ \hookrightarrow^{\pi^+} \circ \hookrightarrow^{\pi^+ *} {\tt S}$, where the $\pi^+$-rule following $\sf J$ duplicates either the upper or the lower $\sf J$-flag. 
\end{proposition}

\begin{proof}
We proceed by induction on $|\mathbf{k}|$ for $\sf J$ applied at $\mathbf{k} \in {\sf Pos}({\tt T})$.

\begin{enumerate}[wide, labelwidth=!, labelindent=0pt]
\item[\emph{Base case} ($\mathbf{k} = \epsilon$)]: By applying Lemma \ref{DuplicationsFirstMinDepth}, we can reorganize the replicative rules as ${\tt T} \hookrightarrow^{\sf J} \circ \hookrightarrow^{\Omega_+^{0}} \circ \hookrightarrow^{\Omega_+^{>0}} {\tt S}$, where the rules in $\Omega_+^0$ are applied at the empty position, and those in $\Omega_+^{>0}$ are applied at non-empty positions. If the upper $\sf J$-flag appears more than once in ${\tt S}$, we choose to start duplicating it, transforming the sequence to ${\tt T} \hookrightarrow^{\sf J} \circ \hookrightarrow^{\pi^+} \circ \hookrightarrow^{\hat{\Omega}+^{0}} \circ \hookrightarrow^{\Omega+^{>0}} \circ \hookrightarrow^{\sigma *} {\tt S}$, thanks to Lemma \ref{DuplicationsAtWidth} and Corollary \ref{CorollarySigma}, which ensures that structural rules are delayed to the end. Otherwise, if no such duplication occurs, the $\sf J$-flag remains unduplicated when applying $\Omega_+^0$. Therefore, by consecutively applying Proposition \ref{ChildDupModal}, we permute the rewrite sequence ${\tt T} \hookrightarrow^{\Omega_+^0} \circ \hookrightarrow^{\sf J} \hat{\tt S} \hookrightarrow^{\Omega_+^{>0}} {\tt S}$. We also assume that the upper $\sf J$-flag is placed at position $i$ in $\hat{\tt S}$. By combining Lemmas \ref{LastDuplicateStarInternalLabel}, \ref{DuplicationsFirstMinDepth}, and \ref{DuplicationsAtWidth}, along with Corollary \ref{CorollarySigma}, we conclude the case by reorganizing $\Omega_+^{>0}$ as ${\tt T} \hookrightarrow^{\Omega_+^0} \circ \hookrightarrow^{\sf J} \circ \hookrightarrow^{\pi^+} \circ \hookrightarrow^{\hat{\Omega}^1_i} \circ \hookrightarrow^{\Omega_i^{> 1}} \circ \hookrightarrow^{\Omega_+^{> 0} \setminus \Omega_i} \circ \hookrightarrow^{\sigma *} {\tt S}$, where $\pi^+$ duplicates the lower $\sf J$-flag, the replicative rules in $\hat{\Omega}i^{1}$ are applied at position $i$, $\Omega_i^{> 1}$ consists of replicative rules applied at positions starting from $i$ with a length greater than 1, and $\Omega+^{> 0} \setminus \Omega_i$ is the sublist of $\Omega_+^{> 0}$ applied at positions not starting with $i$.

\item[\emph{Inductive step} ($\mathbf{k} = l \mathbf{r}$)]: By Lemma \ref{DuplicationsFirstMinDepth}, we can similarly reorganize the replicative rules as ${\tt T} \hookrightarrow^{\sf J} \circ \hookrightarrow^{\Omega_+^{0}} \circ \hookrightarrow^{\Omega_+^{>0}} {\tt S}$. We proceed by cases on the number of occurrences of the internal $\sf J$-flag. 

Suppose that the internal $\sf J$-flag occurs more than once in ${\tt S}$. In this case, using Lemma \ref{DuplicationsAtWidth} and Corollary \ref{CorollarySigma}, we can conclude that ${\tt T} \hookrightarrow^{\sf J} \circ \hookrightarrow^{\pi^+} \circ \hookrightarrow^{\hat{\Omega}_+^{0}} \circ \hookrightarrow^{\Omega_+^{>0}} \circ \hookrightarrow^{\sigma *} {\tt S}$, where the first $\pi^+$-rule duplicates the internal $\sf J$-flag (and consequently also the upper and lower $\sf J$-flags that are within the subtree).

Otherwise, if the internal $\sf J$-flag occurs only once in ${\tt S}$, no $\sf J$-flag is duplicated when applying $\Omega_+^0$. In this case, by consecutively applying Proposition \ref{ChildDupModal}, we obtain ${\tt T} \hookrightarrow^{\Omega_+^0} \circ \hookrightarrow^{\sf J} \bar{\tt S}  \hookrightarrow^{\Omega_+^{>0}} {\tt S}$. Let $j$ denote the position of the internal $\sf J$-flag in $\bar{\tt S}$. Using Lemma \ref{LastDuplicateStarInternalLabel}, we can rewrite this as ${\tt T} \hookrightarrow^{\Omega_+^0} \hat{\tt S} \hookrightarrow^{\sf J} \circ \hookrightarrow^{\Omega_j} \tilde{\tt S} \hookrightarrow^{\Omega_+^{>0} \setminus \Omega_j} {\tt S}$, where $\Omega_j$ is the sublist of $\Omega_+^{>0}$ consisting of rules applied at positions starting by $j$. Then, we have that $\hat{\tt S}|_j \hookrightarrow^{\sf J} \circ \hookrightarrow^{\tilde{\Omega}_j} \tilde{\tt S}|_j$, where $\sf J$ is applied at $\mathbf{r} \in {\sf Pos}(\hat{\tt S}|_j)$ and $\tilde{\Omega}_j = [\pi^+(\mathbf{u},v) \mid \pi^+(j\mathbf{u},v) \in \Omega_j]$. Since the occurrences of the upper and lower $\sf J$-flags remain unchanged within $\tilde{\tt S}|_j$, by the inductive hypothesis we can rewrite $\hat{\tt S}|_j \hookrightarrow^{\pi^+ *} \circ \hookrightarrow^{\sf J} \circ \hookrightarrow^{\pi^+} \circ \hookrightarrow^{\pi^+ *} \tilde{\tt S}|_j$,
ensuring that the $\pi^+$-rule after $\sf J$ duplicates the upper or the lower $\sf J$-flag. By the Deep Rewriting Property (Proposition \ref{InsideRewriting}), we conclude that ${\tt T} \hookrightarrow^{\Omega_+^0} \circ \hookrightarrow^{\pi^+ *} \circ \hookrightarrow^{\sf J} \circ \hookrightarrow^{\pi^+} \circ \hookrightarrow^{\pi^+ *} \circ \hookrightarrow^{\Omega_+^{>0} \setminus \Omega_j} {\tt S}$, where the $\pi^+$-rule following $\sf J$ duplicates the upper or lower $\sf J$-flag. 
\qedhere
\end{enumerate}
\end{proof}

When permuting $\sf J$ with a list of replicative rules, it is crucial to ensure that induction on the number of $\sf J$-flags can proceed. This is guaranteed by showing that the number of $\sf J$-flags introduced during the permutation of $\sf J$ and $\pi^+$ is strictly less than in the original derivation. Figure \ref{fig:ExJLabelsDecrease} illustrates this process by depicting how the flags evolve under such permutations. While new $\sf J$-rules are applied, the number of flags introduced at each step is always lower than in the original derivation. Specifically, in the initial rewriting (a), $\sf J$ is applied at the empty position, with the upper $\sf J$-flag (vertical lines) appearing once and the lower $\sf J$-flag (horizontal lines) occurring multiple times. After permuting the rules in (b), new lower $\sf J$-flags (filled with dots for the first $\sf J$ and with a grid for the second) emerge, but their occurrences are strictly fewer than before. In the following proposition, we formally prove that the number of $\sf J$-flags resulting from the permutation process is always strictly less than in the original derivation.  

\begin{figure}[h]
    \centering
\begin{subfigure}{1\textwidth}
\centering
\begin{subfigure}{.25\textwidth}
    \centering
    \begin{tikzpicture}   
    \node[state][label=right: \hspace{0.4cm}$\hookrightarrow^{\sf J}$] (a) at (0,0) {} ;
    \node[state][pattern = vertical lines] (b) [below left =of a] {};
    \node[state][pattern = horizontal lines] (c) [below right =of a] {};
    \draw[-latex] (a) -- (c) node[fill=white,inner sep=2pt,midway] {};
    \draw[-latex] (a) -- (b) node[fill=white,inner sep=2pt,midway] {};
    \end{tikzpicture} 
\end{subfigure}
\begin{subfigure}{.1\textwidth}
    \centering
    \begin{tikzpicture}
    \node[state] (a) at (0,0) {} ;
    \node[state][pattern = vertical lines][label=right: \hspace{0.4cm}$\hookrightarrow^{\pi^+}$] (b) [below =of a] {};
    \node[state][pattern = horizontal lines] (c) [below =of b] {};
    \draw[-latex] (a) -- (b) node[fill=white,inner sep=2pt,midway] {};
    \draw[-latex] (b) -- (c) node[fill=white,inner sep=2pt,midway] {};
    \end{tikzpicture} 
\end{subfigure}
\begin{subfigure}{.2\textwidth}
    \centering
    \begin{tikzpicture}        
    \node[state] (e) at (0,0) {} ;
    \node[state][pattern = vertical lines][label=right: \hspace{0.9cm}$\hookrightarrow^{\Omega_+}$] (a) [below =of e] {};
    \node[state][pattern = horizontal lines] (b) [below left =of a] {};
    \node[state][pattern = horizontal lines] (c) [below right=of a] {};
    \draw[-latex] (a) -- (b) node[fill=white,inner sep=2pt,midway] {};
    \draw[-latex] (a) -- (c) node[fill=white,inner sep=2pt,midway] {};
    \draw[-latex] (e) -- (a) node[fill=white,inner sep=2pt,midway] {};
    \end{tikzpicture} 
\end{subfigure}
\begin{subfigure}{.2\textwidth}
    \centering
    \begin{tikzpicture}        
    \node[state] (e) at (0,0) {} ;
    \node[state][pattern = vertical lines] (a) [below =of e] {};
    \node[state][pattern = horizontal lines] (b) [below left =of a] {};
    \node[state][pattern = horizontal lines] (c) [below right=of a] {};
    \node[state][pattern = horizontal lines] (d) [below =of a] {};
    \node[state][pattern = horizontal lines] (f) [left =of b] {};
    \node[state][pattern = horizontal lines] (g) [right=of c] {};
    \draw[-latex] (a) -- (b) node[fill=white,inner sep=2pt,midway] {};
    \draw[-latex] (a) -- (c) node[fill=white,inner sep=2pt,midway] {};
    \draw[-latex] (a) -- (d) node[fill=white,inner sep=2pt,midway] {};
    \draw[-latex] (a) -- (f) node[fill=white,inner sep=2pt,midway] {};
    \draw[-latex] (a) -- (g) node[fill=white,inner sep=2pt,midway] {};
    \draw[-latex] (e) -- (a) node[fill=white,inner sep=2pt,midway] {};
    \end{tikzpicture} 
\end{subfigure}
\caption{$\sf J$-flags in the original derivation.}
\end{subfigure}
\begin{subfigure}{1.\textwidth}
\centering
\begin{subfigure}{.15\textwidth}
    \centering
    \begin{tikzpicture}   
    \node[state][label=right: \hspace{0.6cm}$\hookrightarrow^{\pi^+}$] (a) at (0,0) {} ;
    \node[state] (b) [below left =of a] {};
    \node[state] (c) [below right =of a] {};
    \draw[-latex] (a) -- (c) node[fill=white,inner sep=2pt,midway] {};
    \draw[-latex] (a) -- (b) node[fill=white,inner sep=2pt,midway] {};
    \end{tikzpicture} 
\end{subfigure}
\begin{subfigure}{.2\textwidth}
    \centering
    \begin{tikzpicture}
    \node[state][label=right: \hspace{0.3cm}$\hookrightarrow^{\sf J}$] (a) at (0,0) {} ;
    \node[state][pattern = dots] (b) [below left =of a] {};
    \node[state] (c) [below right =of a] {};
    \node[state][pattern = vertical lines] (d) [below =of a] {};
    \draw[-latex] (a) -- (c) node[fill=white,inner sep=2pt,midway] {};
    \draw[-latex] (a) -- (b) node[fill=white,inner sep=2pt,midway] {};
    \draw[-latex] (a) -- (d) node[fill=white,inner sep=2pt,midway] {};
    \end{tikzpicture} 
\end{subfigure}
\begin{subfigure}{.1\textwidth}
    \centering
    \begin{tikzpicture}
    \node[state] (a) at (0,0) {} ;
    \node[state][pattern = vertical lines] (b) [below =of a] {};
    \node[state][pattern = dots] (c) [below =of b] {};
    \node[state][pattern = grid][label=right: \hspace{0.2cm}$\hookrightarrow^{\sf J}$] (d) [below right =of a] {};
    \draw[-latex] (a) -- (b) node[fill=white,inner sep=2pt,midway] {};
    \draw[-latex] (a) -- (d) node[fill=white,inner sep=2pt,midway] {};
    \draw[-latex] (b) -- (c) node[fill=white,inner sep=2pt,midway] {};
    \end{tikzpicture} 
\end{subfigure}
\begin{subfigure}{.1\textwidth}
    \centering
    \begin{tikzpicture}        
    \node[state] (e) at (0,0) {} ;
    \node[state][pattern = vertical lines][label=right: \hspace{0.8cm}$\hookrightarrow^{\Omega_+}$] (a) [below =of e] {};
    \node[state][pattern = dots] (b) [below left =of a] {};
    \node[state][pattern = grid] (c) [below right=of a] {};
    \draw[-latex] (a) -- (b) node[fill=white,inner sep=2pt,midway] {};
    \draw[-latex] (a) -- (c) node[fill=white,inner sep=2pt,midway] {};
    \draw[-latex] (e) -- (a) node[fill=white,inner sep=2pt,midway] {};
    \end{tikzpicture} 
\end{subfigure}
\hspace{0.5cm}
\begin{subfigure}{.3\textwidth}
    \centering
    \begin{tikzpicture}        
    \node[state] (e) at (0,0) {} ;
    \node[state][pattern = vertical lines] (a) [below =of e] {};
    \node[state][pattern = grid] (b) [below left =of a] {};
    \node[state][pattern = dots] (c) [below right=of a] {};
    \node[state][pattern = dots] (d) [below =of a] {};
    \node[state][pattern = dots] (f) [left =of b] {};
    \node[state][pattern = grid] (g) [right=of c] {};
    \draw[-latex] (a) -- (b) node[fill=white,inner sep=2pt,midway] {};
    \draw[-latex] (a) -- (c) node[fill=white,inner sep=2pt,midway] {};
    \draw[-latex] (a) -- (d) node[fill=white,inner sep=2pt,midway] {};
    \draw[-latex] (a) -- (f) node[fill=white,inner sep=2pt,midway] {};
    \draw[-latex] (a) -- (g) node[fill=white,inner sep=2pt,midway] {};
    \draw[-latex] (e) -- (a) node[fill=white,inner sep=2pt,midway] {};
    \end{tikzpicture} 
\end{subfigure}
\caption{$\sf J$-flags after the permutation.}
\end{subfigure}
\caption{Occurrences of the $\sf J$-flags when we permute $\sf J$ and $\pi^+$.}
\label{fig:ExJLabelsDecrease}
\end{figure}

\begin{proposition}
\label{JflagsDecrease}
Let ${\tt T} \hookrightarrow^{\sf J} \circ \hookrightarrow^{\pi^+} \circ \hookrightarrow^{\Omega_+} {\tt S}$, where $\pi^+$ duplicates a $\sf J$-flag, $\Omega_+$ is a list of replicative rules, and $n$ is the total number of upper and lower $\sf J$-flags in ${\tt S}$. Then, we can rewrite ${\tt T} \hookrightarrow^{\pi^+ *} \circ \hookrightarrow^{\sf J} \circ \hookrightarrow^{\sf J} \circ \hookrightarrow^{\Omega_+} \circ \hookrightarrow^{\sigma *} {\tt S}$, where the number of upper and lower $\sf J$-flags in ${\tt S}$ introduced by the first and second $\sf J$-rules, denoted $n_1$ and $n_2$, respectively, satisfy $n_1 \leq n - 1$ and $n_2 \leq n - 1$.
\end{proposition}

\begin{proof}
We proceed by induction on $|\Omega_+|$, starting with the inductive step. 
\begin{enumerate}[wide, labelwidth=!, labelindent=0pt]
\item[\emph{Inductive step}]: Consider the sequence ${\tt T} \hookrightarrow^{\sf J} \circ \hookrightarrow^{\pi^+} \circ \hookrightarrow^{\Omega_+} \tilde{\tt S} \hookrightarrow^{\pi^+} {\tt S}$, where the $\pi^+$-rule following $\sf J$ duplicates a $\sf J$-flag, and let $\tilde{n}$ be the number of upper and lower $\sf J$-flags in $\tilde{\tt S}$. By the inductive hypothesis, we rewrite ${\tt T} \hookrightarrow^{\pi^+ *} \circ \hookrightarrow^{\sf J} \circ \hookrightarrow^{\sf J} \circ \hookrightarrow^{\Omega_+} \circ \hookrightarrow^{\sigma *} \tilde{\tt S} \hookrightarrow^{\pi^+} {\tt S}$ such that $n_1 \leq \tilde{n} - 1$ and $n_2 \leq \tilde{n} - 1$. Therefore, since $\tilde{n} \leq n$, we conclude by applying Corollary \ref{CorollarySigma}.

\item[\emph{Base case}]: Let ${\tt T} \hookrightarrow^{\sf J} \circ \hookrightarrow^{\pi^+} {\tt S}$, where $\pi^+$ duplicates a $\sf J$-flag. We continue by induction on $|\mathbf{k}|$ such that $\sf J$ is applied at $\mathbf{k} \in {\sf Pos}({\tt T})$.
\begin{enumerate}[wide, labelwidth=!, labelindent=10pt]
\item[{\emph Base case} ($\mathbf{k} = \epsilon$)]: In this case, $\pi^+$ duplicates either the upper or the lower $\sf J$-flag. If $\pi^+$ duplicates the upper $\sf J$-flag, then $n = 4$, as shown in Figure \ref{fig:JPi+1Normalization}. We can then rewrite the expression according to Proposition \ref{ChildDupModal} as ${\tt T} \hookrightarrow^{\pi^+} \circ \hookrightarrow^{\pi^+} \circ \hookrightarrow^{\sf J} \circ \hookrightarrow^{\sf J} {\tt S}$, where $n_1 = 2$ and $n_2 = 2$. Otherwise, if the upper $\sf J$-flag occurs only once in ${\tt S}$, then $\pi^+$ duplicates the lower $\sf J$-flag. In this case, $n = 3$, as illustrated in Figure \ref{fig:JPi+2Normalization}. We can rewrite it as ${\tt T} \hookrightarrow^{\pi^+} \circ \hookrightarrow^{\sf J} \circ \hookrightarrow^{\sf J} \circ \hookrightarrow^{\sigma *} {\tt S}$, with $n_1 = 2$ and $n_2 = 2$, just as in the case from Proposition \ref{ChildDupModal}.

\item[\emph{Inductive step} ($\mathbf{k} = i\mathbf{r}$)]: If $\pi^+$ duplicates the internal $\sf J$-flag, then $n = 4$. In this case, by Proposition \ref{ChildDupModal}, we can rewrite the expression as ${\tt T} \hookrightarrow^{\pi^+} \circ \hookrightarrow^{\sf J} \circ \hookrightarrow^{\sf J} {\tt S}$, with $n_1 = 2$ and $n_2 = 2$. On the other hand, if $\pi^+$ is applied within the subtree ${\tt T}|_i$, duplicating either the upper or lower $\sf J$-flag, we have that ${\tt T}|_i \hookrightarrow^{\sf J} \circ \hookrightarrow^{\pi^+} {\tt S}|_i$, where $\sf J$ is applied at position $\mathbf{r} \in {\sf Pos}({\tt T}|_i)$ and $\pi^+$ duplicates the upper or lower $\sf J$-flag. Notably, the number of occurrences of the upper and the lower $\sf J$-flags in ${\tt S}|_i$ is also $n$. Therefore, applying the inductive hypothesis, we can rewrite it as ${\tt T}|_i \hookrightarrow^{\pi^+} \circ \hookrightarrow^{\sf J} \circ \hookrightarrow^{\sf J} \circ \hookrightarrow^{\sigma *} {\tt S}|_i$, where $n_1 \leq n - 1$ and $n_2 \leq n - 1$ represent the number of occurrences of $\sf J$-flags in ${\tt S}|_i$ as determined by the first and second $\sf J$-rules, respectively. Finally, by the Deep Rewriting Property (Proposition \ref{InsideRewriting}), we conclude that ${\tt T} \hookrightarrow^{\pi^+} \circ \hookrightarrow^{\sf J} \circ \hookrightarrow^{\sf J} \circ \hookrightarrow^{\sigma *} {\tt S}$, as the number of occurrences of $\sf J$-flags coincides with $n_1$ and $n_2$ by construction.
\qedhere
\end{enumerate}
\end{enumerate} 
\end{proof}

Thus, we can now show how $\sf J$ permutes with a list of replicative rules.

\begin{proposition}
\label{JSigmaReplicative}
Let ${\tt T} \hookrightarrow^{\sf J} \circ \hookrightarrow^{\Omega_+} {\tt S}$, where $\Omega_+$ is a list of replicative rules and $n$ is the total number of upper and lower $\sf J$-flags occurring in ${\tt S}$. Then, we can rewrite ${\tt T} \hookrightarrow^{\pi^+ *} \circ \hookrightarrow^{\Omega_{\sf J}} \circ \hookrightarrow^{\sigma *} {\tt S}$, where $\Omega_{\sf J}$ is a list of $\sf J$-rules satisfying $|\Omega_{\sf J}| \leq 2^n$.
\end{proposition}

\begin{proof}
We proceed by induction on $n - 2$. The base case is trivial since no $\sf J$-flag is duplicated, so we obtain ${\tt T} \hookrightarrow^{\Omega_+} \circ \hookrightarrow^{\sf J} {\tt S}$ directly by a consecutive application of Proposition \ref{ChildDupModal}. Otherwise, assume ${\tt T} \hookrightarrow^{\sf J} \circ \hookrightarrow^{\Omega_+} {\tt S}$, where the upper and lower $\sf J$-flags occur $n$ times in ${\tt S}$. By Proposition \ref{StartDupJFlag}, we can choose to duplicate either the upper or lower $\sf J$-flag after $\sf J$, rewriting the derivation as ${\tt T} \hookrightarrow^{\pi^+ *} \circ \hookrightarrow^{\sf J} \circ \hookrightarrow^{\pi^+} \circ \hookrightarrow^{\tilde{\Omega}_+} {\tt S}$. Furthermore, by Proposition \ref{JflagsDecrease}, we can rewrite it as ${\tt T} \hookrightarrow^{\pi^+ *} \circ \hookrightarrow^{\sf J} \circ \hookrightarrow^{\sf J} \circ \hookrightarrow^{\tilde{\Omega}_+} \circ \hookrightarrow^{\sigma *} {\tt S}$, ensuring that the occurrences of the upper and lower $\sf J$-flags of each $\sf J$-rule are at most $n - 1$ in ${\tt S}$. Then, applying the inductive hypothesis, we obtain ${\tt T} \hookrightarrow^{\pi^+ *} \circ \hookrightarrow^{\sf J} \circ \hookrightarrow^{\pi^+ *} \circ \hookrightarrow^{\Omega_{\sf J}^{IH1}}  \circ \hookrightarrow^{\sigma *} {\tt S}$, where $|\Omega_{\sf J}^{IH1}| \leq 2^{n - 1}$. Since the occurrences of the remaining $\sf J$-rule stay the same, again by the inductive hypothesis we deduce ${\tt T} \hookrightarrow^{\pi^+ *} \circ \hookrightarrow^{\Omega_{\sf J}^{IH2}} \circ \hookrightarrow^{\sigma *} \circ \hookrightarrow^{\Omega_{\sf J}^{IH1}} \circ \hookrightarrow^{\sigma *} {\tt S}$ with $|\Omega_{\sf J}^{IH2}| \leq 2^{n - 1}$. Finally, using Corollary \ref{CorollarySigma} to move structural rules to the end, we conclude ${\tt T} \hookrightarrow^{\pi^+ *} \circ \hookrightarrow^{\Omega_{\sf J}^{IH2}} \circ  \hookrightarrow^{\Omega_{\sf J}^{IH1}}  \circ \hookrightarrow^{\sigma *} {\tt S}$, where $|\Omega_{\sf J}^{IH1} \smallfrown \Omega_{\sf J}^{IH2}| \leq 2^{n}$.
\end{proof}

Next, we examine how a single $\pi^+$-rule interacts with a list of modal rules. To do so, we first analyze how the $\sf m$-rule behaves with a list of replicative rules.

\begin{lemma}
\label{MonotonicityReplicative}
Let $\Omega_+$ be a list of replicative rules. If ${\tt T} \hookrightarrow^{\sf m} \circ \hookrightarrow^{\Omega_+} {\tt S}$, then ${\tt T} \hookrightarrow^{\Omega_+} \circ \hookrightarrow^{\Omega_{\sf m}} {\tt S}$, where $\Omega_{\sf m}$ is a list of $\sf m$-rules satisfying $|\Omega_{\sf m}| \leq 2^{|\Omega_+|}$.
\end{lemma}

\begin{proof}
In the proof of Lemma \ref{PermutingReplicativeWithList}, we saw that if ${\tt T} \hookrightarrow^{\sf m} \circ \hookrightarrow^{\pi^+} {\tt S}$, then we can rewrite ${\tt T} \hookrightarrow^{\pi^+} \circ \hookrightarrow^{\sf m} {\tt S}$ or ${\tt T} \hookrightarrow^{\pi^+} \circ \hookrightarrow^{\sf m} \circ \hookrightarrow^{\sf m} {\tt S}$. Thus, we conclude by an easy induction on $|\Omega_+|$, leaving the fine details to the reader. 
\end{proof}

Using Proposition \ref{JSigmaReplicative} and Lemma \ref{MonotonicityReplicative}, we can now see how a single application of $\pi^+$ permutes with a list of modal rules.

\begin{corollary}
\label{StarModalStarReplicative}
Let $\Omega_\diamond$ be a list of modal rules. If ${\tt T} \hookrightarrow^{\Omega_\diamond} \circ \hookrightarrow^{\pi^+} {\tt S}$, then ${\tt T} \hookrightarrow^{\Omega_+} \circ \hookrightarrow^{\tilde{\Omega}_\diamond} \circ \hookrightarrow^{\sigma *} {\tt S}$, where $\Omega_+$ and $\tilde{\Omega}_\diamond$ are lists of replicative and modal rules, respectively, satisfying $|\Omega_+| \leq {\sf n}({\tt T})$ and $|\tilde{\Omega}_\diamond| \leq |\Omega_\diamond| \cdot 4^{{\sf n}({\tt T})}$.
\end{corollary}

\begin{proof}
We proceed by induction on $|\Omega_\diamond|$. For the base case, if ${\tt T} \hookrightarrow^{\pi^+} {\tt S}$, the result is trivially satisfied as $1 \leq {\sf n}({\tt T})$. Otherwise, we consider 
\begin{equation}
{\tt T} \hookrightarrow^{\theta} \circ \hookrightarrow^{\Omega_\diamond} \circ \hookrightarrow^{\pi^+} {\tt S}
\label{eq:fromIH}
\end{equation} 
for $\theta \in \{{\sf m}, {\sf J}\}$. By the inductive hypothesis, we can rewrite it as ${\tt T} \hookrightarrow^{\theta} {\tt S}_1 \hookrightarrow^{\Omega_+} \circ \hookrightarrow^{\tilde{\Omega}_\diamond} \circ \hookrightarrow^{\sigma *} {\tt S}$ such that $|\Omega_+| \leq {\sf n}({\tt T})$ and $|\tilde{\Omega}_\diamond| \leq |\Omega_\diamond| \cdot 4^{{\sf n}({\tt T})}$. Note that the number of nodes of ${\tt S}_1$ is equal to the number of nodes of ${\tt T}$. We continue by cases on $\theta$. 
\begin{enumerate}[wide, labelwidth=!, labelindent=0pt]
    \item[($\theta = {\sf m}$)]: Using Lemma \ref{MonotonicityReplicative}, we show that ${\tt T} \hookrightarrow^{\Omega_+} \circ \hookrightarrow^{\Omega_{\sf m}} \circ \hookrightarrow^{\tilde{\Omega}_\diamond} \circ \hookrightarrow^{\sigma *} {\tt S}$ for $\Omega_{\sf m}$ a list of $\sf m$-rules satisfying $|\Omega_{\sf m}| \leq 2^{|\Omega_+|}$. Therefore, we have $|\Omega_{\sf m} \smallfrown \tilde{\Omega}_\diamond| \leq 2^{{\sf n}({\tt T})} + |\Omega_\diamond| \cdot 4^{{\sf n}({\tt T})} \leq |\theta \smallfrown \Omega_\diamond| \cdot 4^{{\sf n}({\tt T})}$.
    \item[($\theta = {\sf J}$)]: By Proposition \ref{JSigmaReplicative}, we can rewrite ${\tt T} \hookrightarrow^{\hat{\Omega}_+} \circ \hookrightarrow^{\Omega_{\sf J}} \circ \hookrightarrow^{\sigma *} \circ \hookrightarrow^{\tilde{\Omega}_\diamond} \circ \hookrightarrow^{\sigma *} {\tt S}$, where $\hat{\Omega}_+$ and $\Omega_{\sf J}$ are lists of replicative and $\sf J$-rules, respectively, and $|\Omega_{\sf J}| \leq 2^{{\sf n}({\tt S})}$. Here, the occurrences of the upper and lower $\sf J$-flags are bounded by the total number of nodes. Thus, we obtain rewriting sequence 
    \begin{equation}
    {\tt T} \hookrightarrow^{\hat{\Omega}_+} \circ \hookrightarrow^{\Omega_{\sf J}} \circ \hookrightarrow^{\tilde{\Omega}_\diamond} \circ \hookrightarrow^{\sigma *} {\tt S}
    \label{eq:fromJstep}
    \end{equation}
    by delaying the structural rules at the end using Corollary \ref{CorollarySigma}. Moreover, from (\ref{eq:fromIH}), using Lemma \ref{TreeMetrics} we deduce that ${\sf n}({\tt S}) \leq 2 \cdot {\sf n}({\tt T})$. Hence, we show that $|\Omega_{\sf J} \smallfrown \tilde{\Omega}_\diamond| \leq 2^{2 \cdot {\sf n}({\tt T})} + |\Omega_\diamond| · 4^{{\sf n}({\tt T})}  = |\theta \smallfrown \Omega_\diamond| · 4^{{\sf n}({\tt T})}$. Furthermore, applying Lemma \ref{StarReplicativeNodes} in (\ref{eq:fromJstep}), we obtain $|\hat{\Omega}_+| \leq {\sf n}({\tt S}) - {\sf n}({\tt T}) \leq 2 \cdot {\sf n}({\tt T}) - {\sf n}({\tt T}) = {\sf n}({\tt T})$.
\qedhere
\end{enumerate}
\end{proof}

\subsection{The Rewrite Normalization theorem}

Thus, we conclude with the main theorem of this section.

\begin{theorem}[Rewrite Normalization]
\label{UBNormalization}
Let ${\bf L}$ be a spi-logic such that ${\bf L} \subseteq \{({\sf 4}), ({\sf m}), ({\sf J})\}$, and let $\Omega$ be a list of rules of ${\sf T}{\bf L}$. If ${\tt T} \hookrightarrow^{\Omega} {\tt S}$, we can equivalently rewrite it as ${\tt T} \hookrightarrow^{\bar{\Omega}} {\tt S}$ for $\bar{\Omega}$ a normal rewriting sequence of ${\sf T}{\bf L}$.  

Let $\Omega_{+}, \Omega_\diamond, \Omega_\delta, \Omega_\rho$ and $\Omega_\sigma$ denote the lists of replicative, modal, decreasing, atomic, and structural rules, respectively, composing $\bar{\Omega}$.
\begin{itemize}[wide, labelwidth=!, labelindent=0pt]
    \item If ${\bf L}$ does not include axiom \textnormal{($\sf J$)}, then:
  \[
  |\Omega_+| \leq |\Omega|, \quad
  |\Omega_\diamond| \leq 2^{|\Omega_\delta| + |\Omega|}, \quad
  |\Omega_{\delta} \smallfrown \Omega_\rho| \leq 2^{|\Omega|}, \quad
  |\Omega_\sigma| \leq {\sf n}({\tt S}) - 1.
  \]
  \item Otherwise,  
  \[
  |\Omega_+| \leq ({\sf w}({\tt T}) +1)^{({\sf h}({\tt T}) + 1)^{2 \cdot |\Omega|}}, \quad
  |\Omega_\diamond| \leq 2^{|\Omega_\delta| + 2 \cdot |\Omega| \cdot (|\Omega_+| + {\sf w}({\tt T}) + 1)^{{\sf h}({\tt T})}},  
  \]
  \[
  |\Omega_\delta \smallfrown \Omega_\rho| \leq 2^{|\Omega|}, \quad
  |\Omega_\sigma| \leq {\sf n}({\tt S}) - 1.
  \]
\end{itemize}
\end{theorem}

\begin{proof}
We proceed by induction on $|\Omega|$, where the base case is trivially satisfied. Otherwise, we consider the rewriting ${\tt T} \hookrightarrow^{\Omega} \hat{\tt S} \hookrightarrow^{\mu} {\tt S}$ for $\hat{\tt S} \in {\sf Tree^\diamond}$ and $\mu$ a rule of ${\sf T}{\bf L}$. By the inductive hypothesis it follows that 
\begin{equation}
{\tt T} \hookrightarrow^{\Omega_{+}} \circ \hookrightarrow^{\Omega_{\diamond}} \circ \hookrightarrow^{\Omega_\delta} \circ \hookrightarrow^{\Omega_\rho} \circ \hookrightarrow^{\Omega_\sigma} \hat{\tt S} \hookrightarrow^{\mu} {\tt S}
\label{eq:initialrewriting}
\end{equation}  
for lists of rules satisfying the bounds. We continue by cases on the kind of $\mu$.

\begin{enumerate}[wide, labelwidth=!, labelindent=0pt]
    \item[($\mu$\emph{ is a structural rule})]: By Proposition \ref{UBAnyPermutation}, we can equivalently rewrite ${\tt T} \hookrightarrow^{\Omega_{+}} \circ \hookrightarrow^{\Omega_{\diamond}} \circ \hookrightarrow^{\Omega_\delta} \circ \hookrightarrow^{\Omega_\rho} \bar{\tt S} \hookrightarrow^{\bar{\Omega}_\sigma} {\tt S}$ for $\bar{\Omega}_\sigma$ a list of structural rules satisfying $|\bar{\Omega}_\sigma| \leq {\sf n}(\bar{\tt S}) - 1 = {\sf }n({\tt S}) - 1$. The remaining bounds are straightforward.

    \item[($\mu$\emph{ is an atomic rule})]: We simply conclude ${\tt T} \hookrightarrow^{\Omega_{+}} \circ \hookrightarrow^{\Omega_{\diamond}} \circ \hookrightarrow^{\Omega_\delta} \circ \hookrightarrow^{\Omega_\rho} \circ \hookrightarrow^{\mu} \circ \hookrightarrow^{\Omega_\sigma} {\tt S}$ by Lemma \ref{PermuteWithStructural}. Hence, the bounds can be easily verified. In particular, $|\Omega_\delta \smallfrown \Omega_\rho \smallfrown \mu| \leq 1 + 2^{|\Omega|} \leq 2^{|\Omega \smallfrown \mu|}$.

    \item[($\mu$\emph{ is a decreasing rule})]: By Lemma \ref{PermuteWithStructural}, we have that ${\tt T} \hookrightarrow^{\Omega_{+}} \circ \hookrightarrow^{\Omega_{\diamond}} \circ \hookrightarrow^{\Omega_\delta} \circ \hookrightarrow^{\Omega_\rho} \tilde{\tt S} \hookrightarrow^{\mu} \circ \hookrightarrow^{\tilde{\Omega}_\sigma} {\tt S}$ for $\tilde{\Omega}_\sigma$ a list of structural rules satisfying $|\tilde{\Omega}_\sigma| \leq |\Omega_\sigma| \cdot {\sf w}(\tilde{\tt S}) \leq |\Omega_\sigma| \cdot {\sf w}({\tt S})$. Using Lemma \ref{PermuteWithAtomic} we can rewrite ${\tt T} \hookrightarrow^{\Omega_{+}} \circ \hookrightarrow^{\Omega_{\diamond}} \circ \hookrightarrow^{\Omega_\delta} \circ \hookrightarrow^{\mu} \circ \hookrightarrow^{\tilde{\Omega}_\rho} \circ \hookrightarrow^{\tilde{\Omega}_\sigma} {\tt S}$, where $\tilde{\Omega}_\rho$ is a list of atomic rules that satisfies $|\tilde{\Omega}_\rho| \leq |\Omega_\rho|$. Moreover, by Proposition \ref{UBAnyPermutation}, we can rewrite ${\tt T} \hookrightarrow^{\Omega_{+}} \circ \hookrightarrow^{\Omega_{\diamond}} \circ \hookrightarrow^{\Omega_\delta} \circ \hookrightarrow^{\mu} \circ \hookrightarrow^{\tilde{\Omega}_\rho} \circ \hookrightarrow^{\bar{\Omega}_\sigma} {\tt S}$ for $\bar{\Omega}_\sigma$ a list of structural rules satisfying $|\bar{\Omega}_\sigma| \leq {\sf n}({\tt S}) - 1$. Then, the bounds are straightforward, in particular $|\Omega_\delta \smallfrown \mu \smallfrown \tilde{\Omega}_\rho| \leq 1 + |\Omega_\delta \smallfrown \Omega_\rho| \leq 2^{|\Omega \smallfrown \mu|}$.

    \item[($\mu$\emph{ is a modal rule})]: If ${\bf L}$ does not include axiom ($\sf J$), using Lemma \ref{PermuteWithStructural}, Lemma \ref{PermuteWithAtomic} and Lemma \ref{PermuteWithDecreasing}, it follows that ${\tt T} \hookrightarrow^{\Omega_{+}} \circ \hookrightarrow^{\Omega_{\diamond}} \circ \hookrightarrow^{\Omega_{\mu}} \circ \hookrightarrow^{\Omega_\delta} \circ \hookrightarrow^{\Omega_\rho} \circ \hookrightarrow^{\Omega_\sigma} {\tt S}$, where $\Omega_\mu$ is a list of $\mu$-rules satisfying $|\Omega_\mu| \leq 2^{|\Omega_\delta|}$. Thus, by the inductive hypothesis we have that $|\Omega_\diamond \smallfrown \Omega_\mu| \leq 2^{|\Omega| + |\Omega_\delta|} + 2^{|\Omega_\delta|} \leq 2^{|\Omega \smallfrown \mu| + |\Omega_\delta|}$. The rest of the bounds are straightforward.
    
    Otherwise, if ${\bf L}$ includes axiom ($\sf J$), applying Lemma \ref{PermuteWithStructural}, Lemma \ref{PermuteWithAtomic} and Lemma \ref{PermuteWithDecreasing} we have that ${\tt T} \hookrightarrow^{\Omega_{+}} \circ \hookrightarrow^{\Omega_{\diamond}} \circ \hookrightarrow^{\Omega_{\mu}} \circ \hookrightarrow^{\Omega_\delta} \circ \hookrightarrow^{\Omega_\rho} \circ \hookrightarrow^{\tilde{\Omega}_\sigma} {\tt S}$, for $\Omega_\mu$ is a list of $\mu$-rules satisfying $|\Omega_\mu| \leq 2^{|\Omega_\delta|}$ and $\tilde{\Omega}_\sigma$ a list of structural rules. Moreover, by Proposition \ref{UBAnyPermutation} we can rewrite ${\tt T} \hookrightarrow^{\Omega_{+}} \circ \hookrightarrow^{\Omega_{\diamond}} \circ \hookrightarrow^{\Omega_{\mu}} \circ \hookrightarrow^{\Omega_\delta} \circ \hookrightarrow^{\Omega_\rho} \circ \hookrightarrow^{\bar{\Omega}_\sigma} {\tt S}$ for $\bar{\Omega}_\sigma$ a list of structural rules satisfying $|\bar{\Omega}_\sigma| \leq {\sf n}({\tt S}) - 1$. Moreover, using the inductive hypothesis we establish that $|\Omega_\diamond \smallfrown \Omega_\mu|  \leq 2^{|\Omega_\delta| + 2 \cdot |\Omega| \cdot (|\Omega_+| + {\sf w}({\tt T}) + 1)^{{\sf h}({\tt T})}} + 2^{|\Omega_\delta|} 
    \leq 2^{|\Omega_\delta| + 2 \cdot |\Omega \smallfrown \mu| \cdot (|\Omega_+| + {\sf w}({\tt T}) + 1)^{{\sf h}({\tt T})}}$. The remaining bounds are straightforward. 

    \item[($\mu$\emph{ is a replicative rule})]: By a consecutive application of Lemma \ref{PermutingReplicativeWithList}, we can rewrite ${\tt T} \hookrightarrow^{\Omega_{+}} \circ \hookrightarrow^{\Omega_{\diamond}} \circ \hookrightarrow^{\mu} \circ \hookrightarrow^{\tilde{\Omega}_\delta} \circ \hookrightarrow^{\tilde{\Omega}_\rho} \circ \hookrightarrow^{\tilde{\Omega}_\sigma} {\tt S}$, where $\tilde{\Omega}_\delta$, $\tilde{\Omega}_\rho$ and $\tilde{\Omega}_\sigma$ are lists of decreasing, atomic and structural rules, respectively, satisfying $|\tilde{\Omega}_\delta| \leq 2 \cdot |\Omega_\delta|$, $|\tilde{\Omega}_\rho| \leq 2 \cdot |\Omega_\rho|$ and $|\tilde{\Omega}_\sigma| \leq 2 \cdot |\Omega_\sigma|$. To permute $\mu$ with $\Omega_\diamond$, we consider cases on wether ${\bf L}$ includes axiom ($\sf J$) or not.
    
    Suppose that ${\bf L}$ does not include axiom ($\sf J$). Then, by Lemma \ref{PermutingReplicativeWithList} it follows that ${\tt T} \hookrightarrow^{\Omega_{+}} \circ 
    \hookrightarrow^{\mu} \circ \hookrightarrow^{\tilde{\Omega}_{\diamond}} \circ \hookrightarrow^{\tilde{\Omega}_\delta} \circ \hookrightarrow^{\tilde{\Omega}_\rho} \circ \hookrightarrow^{\tilde{\Omega}_\sigma} {\tt S}$ for $\tilde{\Omega}_{\diamond}$ a list of $\sf m$-rules satisfying $|\tilde{\Omega}_{\diamond}| \leq 2 \cdot |\Omega_{\diamond}|$. Using Proposition \ref{UBAnyPermutation}, we can rewrite ${\tt T} \hookrightarrow^{\Omega_{+}} \circ 
    \hookrightarrow^{\mu} \circ \hookrightarrow^{\tilde{\Omega}_{\diamond}} \circ \hookrightarrow^{\tilde{\Omega}_\delta} \circ \hookrightarrow^{\tilde{\Omega}_\rho} \circ \hookrightarrow^{\bar{\Omega}_\sigma} {\tt S}$, where $\bar{\Omega}_\sigma$ is a list of structural rules satisfying $|\bar{\Omega}_\sigma| \leq {\sf n}({\tt S}) - 1$. Moreover, by the inductive hypothesis we show $|\Omega_+ \smallfrown \mu| \leq |\Omega \smallfrown \mu|$, $|\tilde{\Omega}_{\diamond}| \leq 2 \cdot 2^{|\Omega| + |\Omega_\delta|} \leq 2^{|\Omega \smallfrown \mu| + |\Omega_\delta|}$ and $|\tilde{\Omega}_\delta \smallfrown \tilde{\Omega}_\rho| \leq 2 \cdot |\Omega_\delta| + 2 \cdot |\Omega_\rho| \leq 2^{|\Omega \smallfrown \mu|}$.
    
    Otherwise, suppose that ${\bf L}$ includes axiom ($\sf J$). By Corollary \ref{StarModalStarReplicative},
        \begin{equation}
        \label{eq:PermuteJPi+}
            {\tt T} \hookrightarrow^{\Omega_{+}} \tilde{\tt S} \hookrightarrow^{\tilde{\Omega}_{+}} \circ  \hookrightarrow^{\tilde{\Omega}_{\diamond}} \circ \hookrightarrow^{\sigma *} \circ \hookrightarrow^{\tilde{\Omega}_\delta} \circ \hookrightarrow^{\tilde{\Omega}_\rho} \circ \hookrightarrow^{\sigma *} {\tt S},
        \end{equation} 
        where $\tilde{\Omega}_{+}$ and $\tilde{\Omega}_{\diamond}$ are lists of replicative and modal rules, respectively, satisfying $|\tilde{\Omega}_+| \leq {\sf n}(\tilde{\tt S})$ and $|\tilde{\Omega}_\diamond| \leq |\Omega_\diamond| · 4^{{\sf n}(\tilde{\tt S})}$. Using Corollary \ref{CorollarySigma}, we can delay structural rules at the of the sequence as ${\tt T} \hookrightarrow^{\Omega_{+}} \circ \hookrightarrow^{\tilde{\Omega}_{+}} \circ  \hookrightarrow^{\tilde{\Omega}_{\diamond}} \circ \hookrightarrow^{\tilde{\Omega}_\delta} \circ \hookrightarrow^{\tilde{\Omega}_\rho} \circ \hookrightarrow^{\sigma *} {\tt S}$. Thus, by Proposition \ref{UBAnyPermutation} we can rewrite it as ${\tt T} \hookrightarrow^{\Omega_{+}} \circ \hookrightarrow^{\tilde{\Omega}_{+}} \circ  \hookrightarrow^{\tilde{\Omega}_{\diamond}} \circ \hookrightarrow^{\tilde{\Omega}_\delta} \circ \hookrightarrow^{\tilde{\Omega}_\rho} \circ \hookrightarrow^{\bar{\Omega}_\sigma} {\tt S}$
        for $\bar{\Omega}_\sigma$ a list of structural rules satisfying $|\bar{\Omega}_\sigma| \leq {\sf n}({\tt S}) - 1$. Moreover, by the inductive hypothesis we see that $|\tilde{\Omega}_\delta \smallfrown \tilde{\Omega}_\rho| \leq 2^{|\Omega \smallfrown \mu|}$. We now provide a more detailed discussion of the remaining bounds.

        Applying Lemma \ref{StarReplicativeNodes} to (\ref{eq:PermuteJPi+}), we see that $|\Omega_+| \leq {\sf n}(\tilde{\tt S}) - {\sf n}({\tt T})$. Then, $|\Omega_+ \smallfrown \tilde{\Omega}_+| \leq {\sf n}(\tilde{\tt S}) - {\sf n}({\tt T}) + {\sf n}(\tilde{\tt S}) \leq 2 \cdot {\sf n}(\tilde{\tt S})$. To continue, we need to find a bound for ${\sf n}(\tilde{\tt S})$ that depends only on the metrics of ${\tt T}$. Remember that, in Lemma \ref{BoundNodesByWH}, we saw that the number of nodes is bounded as ${\sf n}(\tilde{\tt S}) \leq ({\sf w}(\tilde{\tt S}) + 1)^{{\sf h}(\tilde{\tt S})}$. Then, we obtain ${\sf n}(\tilde{\tt S}) \leq (|\Omega_+| + {\sf w}({\tt T}) + 1)^{{\sf h}({\tt T})}$ since ${\sf w}(\tilde{\tt S}) \leq |\Omega_+| + {\sf w}({\tt T})$ and ${\sf h}(\tilde{\tt S}) = {\sf h}({\tt T})$ by Lemma \ref{TreeMetrics}.
        Thus, by the inductive hypothesis, we see using $2 \leq {\sf w}({\tt T}) + 1$ that
        \begin{equation*}
        \begin{split}
          |\Omega_+ \smallfrown \tilde{\Omega}_+| & \leq 2 \cdot {\sf n}(\tilde{\tt S}) \leq 2 \cdot (|\Omega_+| + {\sf w}({\tt T}) + 1)^{{\sf h}({\tt T})} \leq|\Omega_+|^{{\sf h}({\tt T})} \cdot ({\sf w}({\tt T}) + 1)^{{\sf h}({\tt T}) + 1} \\
          & \leq ({\sf w}({\tt T}) +1)^{({\sf h}({\tt T}) + 1)^{2 \cdot |\Omega| + 1}} \cdot ({\sf w}({\tt T}) + 1)^{{\sf h}({\tt T}) + 1} \leq ({\sf w}({\tt T}) +1)^{({\sf h}({\tt T}) + 1)^{2 \cdot |\Omega \smallfrown \mu|}}.
        \end{split}
        \end{equation*}
        Likewise, by the the inductive hypothesis for the modal rules,
        \begin{equation*}
        \begin{split}
        |\tilde{\Omega}_\diamond| & \leq |\Omega_\diamond| · 4^{{\sf n}(\tilde{\tt S})} \leq 2^{|\Omega_\delta| + 2 \cdot |\Omega| \cdot (|\Omega_+| + {\sf w}({\tt T}) + 1)^{{\sf h}({\tt T})}} \cdot 2^{2 \cdot (|\Omega_+| + {\sf w}({\tt T}) + 1)^{{\sf h}({\tt T})}} \\ & \leq 2^{|\Omega_\delta| + 2 \cdot |\Omega \smallfrown \mu| \cdot (|\Omega_+| + {\sf w}({\tt T}) + 1)^{{\sf h}({\tt T})}}. \qedhere
        \end{split}
        \end{equation*}
\end{enumerate}
\end{proof}

\section{Conclusions and future work}

We have developed a method for designing tree rewriting calculi for spi-logics, demonstrating their effectiveness as provability tools. In particular, our approach ensures that logical inference in the standard logics $\bf K^+$, $\bf K4^+$ and $\bf RC$ can be faithfully captured through tree transformations, grounded in a correspondence between formulas in $\mathscr{L}^+$ and the set of modal trees. A key feature of our framework is that provability in the spi-logics under consideration can be achieved solely through normalized rewriting sequences. Specifically, this means that proving a theorem in $\bf K^+$, $\bf K4^+$ and $\bf RC$ reduces to finding an appropriate normal rewriting sequence for their tree rewriting calculi.

Beyond the considered logics, our framework has the potential to be extended to other positive fragments of (poly)modal logics. In particular, one could consider extending $\bf K^+$ with the following axioms:

\begin{description}
    \item \textbf{($\sf 5$)} $\langle \alpha \rangle \varphi \wedge \langle \alpha \rangle \psi \vdash \langle \alpha \rangle (\varphi \wedge \langle \alpha \rangle \psi)$;
    \item \textbf{($\sf T$)} $\varphi \vdash \langle \alpha \rangle \varphi$.
\end{description} 

One can observe that axiom ($\sf 5$) bears a structural similarity to $\sf J$, differing only in that it applies to equal modalities. Consequently, its rewriting rule would be defined analogously to the $\sf J$-rule, with the modification that the labels of the edges. Moreover, since the interplay between the $\pi^+$-rule and $\sf J$-rule does not depend on the labels of edges, a similar result should hold for the $\sf 5$-rule.

Conversely, the definition of a suitable rule for axiom ($\sf T$) seems to be more challenging. We conjecture that its rewriting rule should be defined by duplicating the tree and attaching the copy as a fresh child. Hence, this approach ensures that information of the root is preserved in the tree. Since this transformation involves a form of structural duplication, we expect complications when attempting to permute this rule with $\sf J$.

\hypersetup{urlcolor=ACMBlue}
Looking forward, our abstract rewriting framework opens avenues for deeper investigations into fundamental properties such as the subformula property and the admissibility of rules, as well as providing a strong foundation for computationally efficient implementations (cf. \cite{wos1992automated, goubault2001proof, newborn2000automated}). Crucially, our work is grounded in a type-theoretic presentation of trees, which serves a dual purpose: (i) enabling a precise and succinct specification of rules, and (ii) making the approach particularly well-suited for formalization in proof assistants. Many of our results have already been implemented \cite{rocq2024} in the Rocq proof assistant, and our ultimate goal is to achieve a full formalization of this work. Given the complexity of normalization theorems and proof-theoretic investigations, where numerous intricate cases must be checked in detail, formalization is not just beneficial but essential. The community has been increasingly adopting formalized proofs (see, e.g.,~\cite{ShillitoG22}). Within the framework of Reflection Calculus, formalization has an added significance: results must not only be true but also provable within suitable systems of arithmetic. Here, formal methods provide unmatched clarity.

Finally, our integration with Rocq offers an additional long-term advantage: proofs verified in Rocq can be automatically extracted into fully verified algorithms, paving the way for provability reasoners with the highest level of reliability attainable with current technology.

\end{document}